\documentclass[11pt]{article}

\usepackage[final]{acl}

\usepackage{times}
\usepackage{latexsym}
\usepackage[table]{xcolor}

\usepackage[T1]{fontenc}

\usepackage[utf8]{inputenc}

\usepackage{microtype}

\usepackage{inconsolata}

\usepackage{graphicx}

\title{Replicating Human Motivated Reasoning Studies with LLMs}

\author{Neeley Pate \\
  University of Rochester \\
  \texttt{npate@ur.rochester.edu} \\\And
Adiba Mahbub Proma \\
  University of Rochester \\
  \\\AND
Hangfeng he \\
  University of Rochester \\
\\\And
James N. Druckman \\
  University of Rochester \\
\\\AND
Daniel C. Molden \\
  Northwestern University \\
\\\And
Gourab Ghoshal \\
  University of Rochester \\
\\\And
Ehsan Hoque \\
  University of Rochester \\}

\begin{document}
\maketitle
\begin{abstract}
Motivated reasoning -- the idea that individuals processing information may be motivated to either arrive at accurate beliefs or arrive at desired conclusions -- has been well-explored as a human phenomenon. However, it remains unclear whether base LLMs are affected by motivational manipulations. Replicating 4 prior political motivated reasoning studies, we find that base LLM behavior does not align with expected human behavior. Furthermore, base LLM behavior across models shares some similarities, such as when selecting to abstain from question answering and incorporating provided arguments into opinions. The results suggest that base LLMs may not emulate human motivated reasoning processes. We emphasize the importance of these findings for researchers using LLMs to for certain tasks such as opinion replication and argument assessment.

\end{abstract}

\section{Introduction}

Motivated reasoning, a concept that describes how individuals process information \cite{Druckman_McGrath_2019}, is often portrayed as involving either the pursuit of a directional (predetermined) conclusion or the pursuit of an accurate (non-directional) conclusion \cite{Kunda_1990}. For example, individuals endorse a policy proposed by their party because they desire holding positions that align with their party leaders, regardless of the arguments for or against the policy (a directional goal). Alternatively, they may instead consider the strength of the arguments for the policy in an effort to form an accurate opinion. A particular motivation can ultimately affect individuals' opinions and actions \cite{Bolsen_Druckman_Cook_2014, Bolsen_Druckman_2015, Mullinix_2016}.

\begin{figure}
    \centering
    \includegraphics[width=0.7\linewidth]{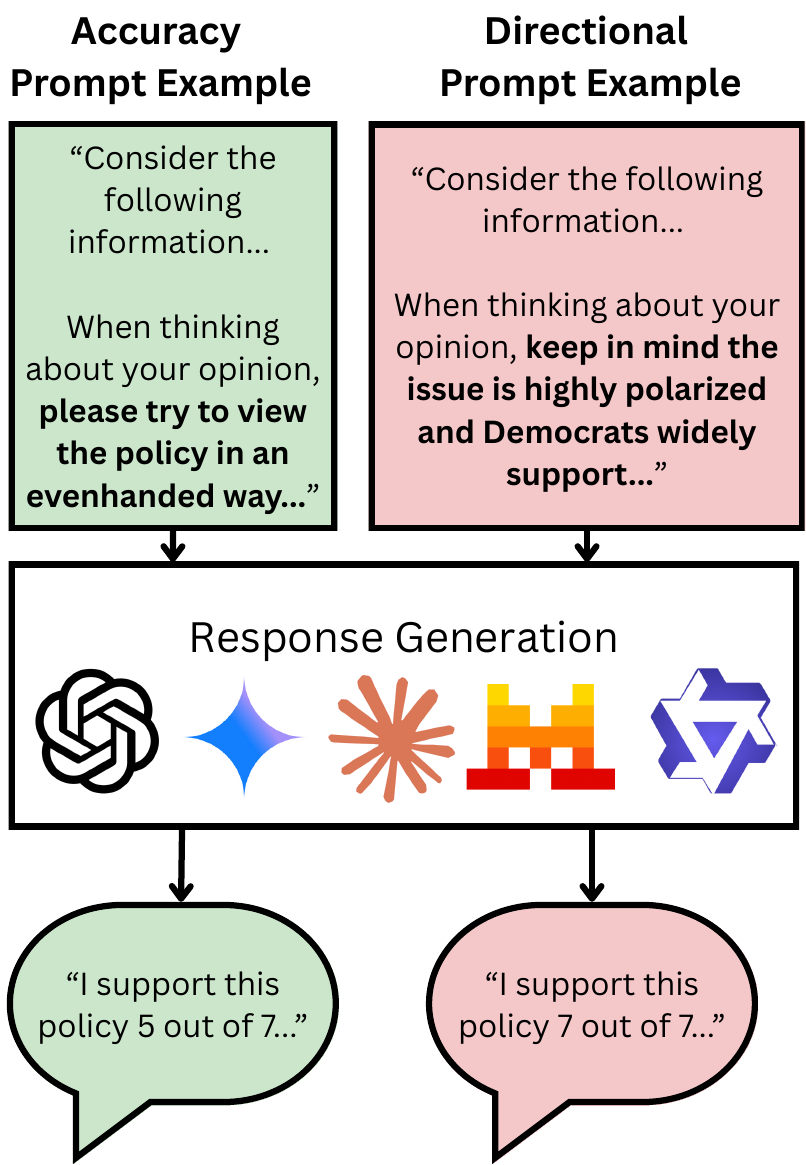}
    \caption{Graphic Describing the Procedure Followed in the Study. An example of different types of prompts (accuracy vs. directional) is shown to highlight potential differences in reasoning.}
    \label{fig:teaser}
    \vspace{-0.5em}
\end{figure}

Motivations affect an individual's opinion formation through how they process information about policies, issues, and stances, particularly within the political domain. The manner in which such information is processed can affect both opinion formation and perceptions of how strong presented arguments are \cite{Druckman_Peterson_Slothuus_2013}. When motivated by accuracy, individuals assess information in a relatively unbiased fashion, focusing on the perceived argument quality rather than external factors such as where their party stands or their existing beliefs \cite{Druckman_McGrath_2019}. When motivated directionally, an individual interprets the information in a way to align with prior beliefs or group expectations -- such as taking a stance consistent with their co-partisans simply because their co-partisans take the position. 


Recent work has begun to explore whether LLMs embody motivated reasoning when given particular aspects of demographics or personality. Studies have shown that providing LLMs with traits such as political affiliation and gender \cite{Dash_Reymond_Spiro_Caliskan_2025} or occupation and goal \cite{Cho_Hoyle_Hermstrwer_2025} affect task outcomes. However, prior work has yet to measure the direct effects of manipulating motivational cues without such personality inductions. To isolate the effects of motivational cues without personality effects, we focus on examining base LLM behaviors, supported by additional testing of prompting for reasoning (Appendix (Appx.) \ref{sec:justification}) and with demographics (Appx. \ref{sec:demographictest}). Thus, this paper aims to answer the following research questions:



\begin{itemize}
    \item \textbf{RQ1}: Do base LLMs (without persona induction) mimic motivated reasoning similar to humans?
    \vspace{-0.5em}
    \item \textbf{RQ2}: Do similarities in behavior exist between groups of LLMs under different motivations and topic domains?
\end{itemize}

In order to answer these questions, we completed a comprehensive literature review to select four prior works on political motivated reasoning to replicate with LLMs (Figure \ref{fig:teaser}). We report results from ten models spanning five families. The contributions of this work include:
\begin{itemize}
    \item Base LLMs do not appear to mimic human-like motivated reasoning.
    \vspace{-0.5em}



    \item When provided with additional pros and cons of policies, LLMs assessments of argument strength and opinions may be misaligned.
    
    \vspace{-0.5em}

    \item Information manipulations may affect when models ``opt-out'' of responding (i.e., do not offer an opinion in response to information), limiting their proxy capabilities in certain settings.
\end{itemize}

These results have implications for social scientists and computer scientists who utilize LLMs as survey proxies or predicting public opinion. Additionally, this work contributes to the artificial intelligence (AI) and natural language processing (NLP) communities for LLM-human alignment or using LLMs for tasks such as argument assessment and opinion formation. We offer considerations for researchers, as motivated reasoning may be one aspect of human-likeness LLMs still struggle to emulate. We also highlight potential pitfalls of LLMs, including unbalanced ``opting out'' of responding and inconsistent strength assessments which could affect LLM judge outcomes.





\section{Related Works}

\subsection{Motivated Reasoning}


Motivated reasoning emphasizes two main avenues of reaching conclusions: pursuing an accuracy (non-directional) goal or pursuing a directional (predetermined) goal \cite{Kunda_1990}. Reasoning is often induced by altering content around or within information. For example, accuracy motivation is often primed by asking participants to consider information provided in an evenhanded fashion \cite{Bolsen_Druckman_Cook_2014, Bolsen_Druckman_2015} or informing participants they will be asked to later justify opinions \cite{Bolsen_Druckman_Cook_2014, Bolsen_Druckman_2015}. To induce directional motivation, there are two broad categories such inductions can fall into: authority/community and personality. Calls on authority or community often include information about who approves (or disapproves), such as party leaders \cite{Bolsen_Druckman_Cook_2014, Druckman_Peterson_Slothuus_2013} or scientific experts \cite{Bolsen_Druckman_2015}. For example, prompting individuals to justify their party preference will make group identity more salient, pushing individuals to align their opinion with the perceived opinion of their party leaders, while not including such information makes these tendencies less prevalent \cite{Bolsen_Druckman_Cook_2014}. Personality techniques call on the individual to reflect on particular cues, such as their prior beliefs and core values \cite{Bayes_Druckman_Goods_Molden_2020}. These cues may bias recall of certain events in a given direction \cite{Kunda_1990}. Pursuing motivated reasoning can affect opinion formation \cite{Bolsen_Druckman_Cook_2014}, argument strength assessments (or how effective arguments for and against the topic were to the person) \cite{Druckman_Peterson_Slothuus_2013}, and statistic interpretation \cite{Kahan_Peters_Dawson_Slovic_2013}. We expand research on motivated reasoning by exploring whether LLMs mimic motivated reasoning from these manipulations.



\subsection{Bias in Persona-Based and Non-Persona-Based LLMs}

Extensive research has explored how biases emerge in LLM responses. It is well-known that LLMs tend to exhibit a left-leaning bias \cite{Bernardelle_Fröhling_Civelli_Lunardi_Roitero_Demartini_2025, Bang_Chen_Lee_Fung_2024, faulborn-etal-2025-little}. Such bias implicit in LLM responses can present in a variety of ways, such as a bias towards certain media outlets \cite{Dai_Cao_Wang_Pang_Xu_Ng_Chua_2025, Proma_Pate_Druckman_Ghoshal_He_Hoque_2025}. Additionally, such biases may be hidden by failing to respond to question answering \cite{Himelstein_LeVi_Youngmann_Nemcovsky_Mendelson_2026}, though work has also emphasized the benefits to user safety of response refusal \cite{Cao_2024}. LLMs have also been known to take on other human-like preferences, such as message structure \cite{li-etal-2025-generation} and self-preference \cite{1Panickssery}. The inclusion of personality traits (or personas) in LLM prompts can also affect how bias presents in LLM responses \cite{Borah_Mihalcea_2024}, though LLMs may fail to reconcile conflicts within a provided persona \cite{Liu_Diab_Fried_2024}. The inclusion of personalities in prompts has been shown to bias LLM behavior in ways similar to motivated reasoning \cite{Cho_Hoyle_Hermstrwer_2025, Dash_Reymond_Spiro_Caliskan_2025}, though they may struggle in identifying human-like motivated reasoning in text \cite{yong-etal-2025-motivebench}. Given prior work's emphasis on the effect of personas on LLM responses, we focus on the whether information, which causes motivated processing in humans, effects LLM response (or non-response), which can inform researchers whether LLMs can be used to emulate or predict human motivated reasoning.

\section{Human Study Selection}
We reviewed literature on political motivated reasoning to choose studies that test for the impact of different motivations on information evaluation and opinion formation.  Selected studies could be completed by providing information in one-shot, and did not require providing the LLM with any personality (i.e., only information relevant to the questions asked were included in the prompt). We also focused on studies that asked similar questions for cross-study analysis. All studies attempt to manipulate motivation, although in varying ways as detailed in Appx. \ref{sec:studydetails} and Table \ref{tab:studyoverview}. In total, we selected four works for replication. A summary of those studies is provided below.

\subsection{Study 1: The Influence of Partisan Motivated Reasoning on Public Opinion}
Study 1 \cite{Bolsen_Druckman_Cook_2014} focuses on how people react to a provided summary of the 2007 Energy Independence Act – whether they oppose or support it (on a 1 to 7 scale). One set of conditions prompted directional goals aimed to motivate people to form opinions consistent with their party elites (about which they learned) by telling participants they would have to justify their party affiliation later on. Another set of conditions prompted an accuracy goal by asking respondents to justify their opinions; a third included no motivation cues at all. There are 15 conditions total, with additional details available in Appx. \ref{sec:studydetails}. The authors find that the directional motivations lead to more agreement with their party relative to the control condition. The accuracy prompts, in contrast, did not alter opinions relative to the control. The evaluative metric was percentage opinion change when provided with a directional or accuracy goal relative to the control condition (Figure \ref{fig:1.1_percentchangepercondition.pdf} ``Human'' panel in Appx. \ref{sec:addres}). 


\subsection{Study 2: Counteracting the Politicization of Science}

Study 2 \cite{Bolsen_Druckman_2015} focuses on whether participants oppose or support (on a 1 to 7 scale) fracking or carbon nanotubes (CNTs) when provided with information on the technologies. Politicized scientific information was provided to lead to directional reasoning; corrections, warnings, and accuracy prompts were conditions meant to lead to accuracy. There are 12 conditions total, with additional details in Appx. \ref{sec:studydetails}. The study finds that exposure to scientific information can boost support while exposure to politicized information can decrease support when compared to the control. Warnings against politicized information, corrections, and accuracy prompts are all effective methods to boost support when encountering politicized information (presented as a percentages of all scores 5 or higher) in each condition (Figure \ref{fig:1.2_percentchangepercondition.pdf} ``Human'' panel in Appx. \ref{sec:addres}).


\subsection{Study 3: How Elite Partisan Polarization Affects Public Opinion Formation}

Study 3 \cite{Druckman_Peterson_Slothuus_2013} investigates how participants respond to context and arguments about drilling for oil and the DREAM Act (Development, Relief, and Education for Alien Minors Act) -- whether they oppose or support it (on a 1 to 7 scale) and perceived argument strength (on a 1 to 7 scale). Directional goals are tied to heightened polarization of an issue; accuracy goals are tied to no polarization around the issue. The authors additionally alter the strength of the arguments provided. This study has 26 conditions, with additional details available in Appx. \ref{sec:studydetails}. The authors find participants polarize in the direction of their party's stance for both opinion formation and argument assessment, but only under the ``polarized'' conditions. When evaluating results, all treatment conditions for opinion formation are compared to a control condition (Figure \ref{fig:1.3_percentchangepercondition.pdf} ``Human'' panels in Appx. \ref{sec:addres}). All treatment conditions for argument assessment evaluate the pro - con ratings within the given condition (Figure \ref{fig:1.3_provcon_permodel.pdf} ``Human'' panels in Appx. \ref{sec:addres}).


\subsection{Study 4: Partisanship and Preference Formation: Competing Motivations, Elite Polarization, and Issue Importance}

Study 4 \cite{Mullinix_2016} focuses on how participants respond to overviews and arguments about either tax cut proposals or the ``Student Success Act'' (SSA) -- whether they oppose or support it (on a 0 to 10 scale) and perceived argument strength (on a 1 to 7 scale). The study invokes directional goals by manipulating polarization and personal importance (whether or not the participant is directly impacted by the proposal). Accuracy goals are tied to no polarization and no importance cues. There are 20 conditions total, with two controls (additional details provided in Appx. \ref{sec:studydetails}). The study finds that party endorsements, issue importance, and polarization play a role in determining issue support and argument strength. We utilize the replication data  \cite{Mullinix_2018}  for the study to report exact measures of means and differences (Figures \ref{fig:1.4_averagesupportpercondition.pdf} and \ref{fig:1.4_provcon_permodel.pdf} ``Human'' panels in Appx. \ref{sec:addres}). 


\section{Experiments and Evaluation}

\subsection{Experimental Setup}

\subsubsection{LLM Selection}

We first surveyed prior work in the domains of politics, psychology, or computer science for human simulation. For each model family selected, we chose one ``reasoning'' and ``non-reasoning'' model to compare. Prior works have used families such as OpenAI and Mistral for modeling capabilities of motivated reasoning under personas \cite{Dash_Reymond_Spiro_Caliskan_2025, yong-etal-2025-motivebench, Cho_Hoyle_Hermstrwer_2025}. Other works have utilized Anthropic \cite{Xie_Feng_Zhang_Li_Yang_Zhang_Feng_He_Gao_Zhang_2024}, Qwen \cite{Liang_Yang_Wang_Xia_Meng_Xu_Wang_Payani_Shu_2025}, and Gemini \cite{Choi_Kim_2025} models for human simulation. Several other models were considered but dropped from the study (details provided in Appx. \ref{sec:llmparam}). The models considered in this study are the following: \textbf{OpenAI}: GPT-4o mini \cite{GPT4omini} and o3-mini \cite{OpenAIo3mini}; \textbf{Google}: Gemini 2.0 Flash \cite{IntroducingGemini2.0_2024} and Gemini 2.5 Flash \cite{Comanici_Bieber_Schaekermann_Pasupat_Sachdeva_Dhillon_Blistein_Ram_Zhang_Rosen_etal_2025}; \textbf{Anthropic}: Claude 3 Haiku \cite{Claude3} and Claude 4.5 Haiku \cite{claude45haiku}; \textbf{Mistral}: Mistral 8x7B  \cite{mistralai} and Mistral Small 4 \cite{mistral4small}; \textbf{Qwen}: Qwen2.5-7B-Instruct \cite{qwen2.5} and Qwen3-8B \cite{qwen3technicalreport}.

The parameters of the models were set to their reported defaults when provided; more information included in Appx. \ref{sec:llmparam}.

\subsubsection{Human Study Replication}
In order to replicate these four studies, we provided the LLMs with the information, including motivation manipulations, and questions asked in the original studies. Models were asked to format their responses by <question number>:<numeric response>. In addition to this condition, a subset of models were tested with a prompt asking to provide reasoning for responses on all studies, discussed in Appx. \ref{sec:justification} and a subset of models were tested with demographics prompts on Study 2, discussed in Appx. \ref{sec:demographictest}. Additional testing supports the findings in the main paper.


In total, the four studies discussed above enlisted 8,558 participants. Rather than creating one-to-one replications of each study, we chose to downsample the number of responses needed. We selected the smallest number of participants for a given condition from all studies (44 \cite{Druckman_Peterson_Slothuus_2013}) to be the smallest number of samples to generate for every study. Setting the minimum condition to 44 generations, the other conditions were adjusted proportionally based on the original sample size of each condition in the original study (e.g., if a study X had 2 conditions, with 400 and 800 people respectively, our generated version would contain 44 and 88 generations). Further justification is provided in Appx. \ref{sec:numsample}. This gives us 3,574 generations across all studies, with the exact number reported in Appx. \ref{sec:numsample}. We replicate every condition tested in the original studies.

\subsection{Evaluation Setup}

For each study, we assess the model's opposition or support for the policy being studied. Additionally, we analyze argument strength when applicable (studies 3 and 4 only). When studies reported results along party lines, we compare the LLM responses to responses by Democratic respondents, given the established left-leaning bias of most models \cite{westwood2025measuring} suggests LLMs will be more likely to respond in ways similar to a left-leaning (Democratic) cohort. Each condition within a study represents a unique combination of information and motivation induction given to participants. When models could not generate a sufficient amount of responses for a given condition (missing > 10\% of the expected generations), those models were dropped from those analyses (denoted by N/A; see Figures \ref{fig:study1_missing.pdf}, \ref{fig:study2_missing.pdf}, \ref{fig:study3_missing_support.pdf},\ref{fig:study3_missing_arg.pdf}, \ref{fig:study4_missing_support.pdf}, and \ref{fig:study4_missing_arg.pdf}.). Aggregate results are reported within the main paper; additional supporting results from are provided in Appx. \ref{sec:addres}. Below outlines the procedure for assessing LLM motivated reasoning.

\subsubsection{Calculating LLM Averages}
For each condition, we compute the average support of each condition for all models. These averages are either on a scale of 1 to 7 (studies 1, 2, and 3) or on a scale of 0 to 10 (study 4). Additionally, we compute the average pro argument assessment and average con argument assessment (on a scale from 1 to 7 for both studies 3 and 4).

\subsubsection{Comparing Human and LLM Response Trends}
We are most interested in understanding if LLMs are capable of embodying human-like motivated reasoning. We investigate this behavior in several ways. First, we compute the Kendall's Tau, a correlation measure, between the average human results and the average LLM results for all conditions, which has been used in prior work \cite{Deutsch_Dror_Roth_2022}. This measure is selected because of tie reconciliation. For our analyses, only correlations that are significant and positive would reflect similar behavior between an LLM and human responses. When discussing results, we match the presentation of those results from the original papers (percent change compared to control mean for study 1 and study 3, percent of responses greater than 4 for study 2, and the mean support for study 4). 


For studies 1 and 3, we also compute the accuracy of the signs of the reported human averages and the calculated LLM averages for all conditions (because we are comparing a change compared to the control condition). When a given mean is reported as less than 0 and significant, it is categorized as -1; likewise positive means that are significant are categorized as 1; all non-significant means are categorized as 0. Significance level is set at p \textless 0.05. 

For studies that evaluate argument strength (studies 3 and 4), we also compute each LLM's ability to assess this. Specifically, we calculate the Kendall's Tau and the accuracy of the sign on the difference in the average pro argument and average con argument assessments for each condition.

\subsubsection{Comparing Different Response Trends Across LLMs}
After looking at the human to model trends, we are interested in identifying trends that all or most models appear to embody. We utilize some of the same techniques for comparing the human response to the LLM response (Kendall's Tau correlation of averages and accuracy of expected sign). We additionally compare the average support and the average argument assessment (pro - con) with Kendall's Tau. This is meant to capture internal LLM assessment alignment (e.g., when pro is stronger than con, the LLM opinion should increase rather than decrease). We also explore what factors cause LLM's to not respond to the support questions, utilizing the Chi-Square measure on the normalized number of missing responses. We report the findings of the argument assessment and non-response in the main paper and report the nuances (mostly model and domain specific) in Appx. \ref{sec:addres} Section \ref{sec:modelspecific}.

\section{Results and Analysis}

\subsection{LLM Behavior Deviates from Human Motivated Reasoning}
\label{sec:RQ1}

\begin{table*}[ht]
  \centering
  \begin{tabular}{|p{0.25\linewidth}|p{0.08\linewidth}|p{0.08\linewidth}|p{0.1\linewidth}|p{0.08\linewidth}||p{0.21\linewidth}|}
    \hline
    \textbf{Model} & \textbf{Study 1} & \textbf{Study 2} & \textbf{Study 3} & \textbf{Study 4} & \textbf{Average Correlation}\\
    \hline
    \textbf{GPT-4o mini} & \cellcolor{yellow!30}$0.529^{*}$& \cellcolor{yellow!30}0.499 & \cellcolor{red!30}-0.258 & \cellcolor{gray!30}0.189 & \cellcolor{gray!30}0.24\\
    \textbf{o3-mini} & \cellcolor{gray!30}0.297 & \cellcolor{red!30}-0.021 & \cellcolor{gray!30}0.25 & \cellcolor{gray!30}0.16 & \cellcolor{gray!30}0.172\\
    \textbf{Gemini 2.0 Flash} & \cellcolor{red!30}-0.029 & N/A & N/A & \cellcolor{red!30}-0.021 & \cellcolor{red!30}-0.025\\
    \textbf{Gemini 2.5 Flash} & N/A & N/A & N/A & \cellcolor{gray!30}0.128& \cellcolor{gray!30}0.128\\
    \textbf{Claude 3 Haiku} & N/A & N/A & \cellcolor{red!30}-0.2 & \cellcolor{gray!30}0.032 & \cellcolor{red!30}-0.084\\
    \textbf{Claude 4.5 Haiku} & N/A & N/A & N/A & N/A & N/A \\
    \textbf{Mistral 8x7B} & \cellcolor{yellow!30}$0.424^{*}$ & N/A & N/A & N/A & \cellcolor{gray!30}$0.424$ \\
    \textbf{Mistral 4 Small} & \cellcolor{gray!30}0.297 & N/A & N/A & \cellcolor{gray!30}0.22 & \cellcolor{gray!30}0.259 \\
    \textbf{Qwen2.5-7B-Instruct} & \cellcolor{gray!30}0.201 & \cellcolor{gray!30}0.306 & \cellcolor{gray!30}$0.314^{*}$ & \cellcolor{gray!30}0.075 &  \cellcolor{gray!30}0.224\\
    \textbf{Qwen3-8B} & \cellcolor{gray!30}0.253 & \cellcolor{gray!30}0.277& N/A & N/A & \cellcolor{gray!30}0.265 \\ \hline
  \end{tabular}
  \caption{Correlations Between Human and LLM Averages. Correlations are computed with Kendall's Tau on the reported average (Study 4), change in average compared to control (Studies 1 and 3), or reported overall support (Study 2). N/A signifies not enough samples were generated in at least one condition to make a fair comparison. * = p value \textless 0.05. Red signifies negative relationships, gray signifies weak positive relationships, and yellow signifies moderate positive relationships.}
  \label{tab:humanvllm}
\end{table*}

\begin{table}[ht]
  \centering
  \begin{tabular}{|p{0.48\linewidth}|p{0.2\linewidth}|p{0.2\linewidth}|}
    \hline
    \textbf{Model} & \textbf{Study 1} & \textbf{Study 3} \\
    \hline
    \textbf{GPT-4o mini} & \cellcolor{yellow!30}0.643 & \cellcolor{gray!30}0.167  \\
    \textbf{o3-mini } & \cellcolor{yellow!30}0.5 & \cellcolor{gray!30}0.542 \\
    \textbf{Gemini 2.0 Flash} &\cellcolor{yellow!30} 0.5 & N/A \\
    \textbf{Gemini 2.5 Flash} & N/A & N/A \\
    \textbf{Claude 3 Haiku} & N/A & \cellcolor{gray!30}0.25 \\
   \textbf{Claude 4.5 Haiku} & N/A & N/A \\
   \textbf{Mistral 8x7B} & \cellcolor{green!30}0.714 & N/A \\
   \textbf{Mistral 4 Small} & \cellcolor{yellow!30}0.571 & N/A \\
    \textbf{Qwen2.5-7B-Instruct} & \cellcolor{gray!30}0.357 & \cellcolor{gray!30}0.25 \\
    \textbf{Qwen3-8B} & \cellcolor{yellow!30}0.5 & N/A \\ 
    \hline
  \end{tabular}
  \caption{Accuracy of the Sign of the Change in Opinion Relative to the Control. N/A signifies not enough samples were generated in at least one condition to make a fair comparison. Gray signifies weak accuracy, yellow signifies moderate accuracy, and green signifies high accuracy.}
  
  \label{tab:accuracysign}
\vspace{-1.5em}
  
\end{table}
  
\begin{table*}
  \centering
  \begin{tabular}{|c|c|c||c|c|}
    \hline
    \textbf{Model} & \textbf{Study 3 Avg.} & \textbf{Study 4 Avg.} & \textbf{Study 3 Sign} & \textbf{Study 4 Sign} \\
    \hline
    \textbf{GPT-4o mini} & \cellcolor{gray!30}$0.383^{**}$ & \cellcolor{red!30}-0.026 & \cellcolor{gray!30}0.458 & \cellcolor{gray!30}0.45 \\
    \textbf{o3-mini} & \cellcolor{gray!30}$0.393^{**}$ & \cellcolor{gray!30}0.18 & \cellcolor{gray!30}0.417 & \cellcolor{gray!30}0.4 \\
    \textbf{Gemini 2.0 Flash} & \cellcolor{gray!30}0.225 & \cellcolor{gray!30}0.095 & \cellcolor{gray!30}0.417 & \cellcolor{yellow!30}0.5 \\
    \textbf{Gemini 2.5 Flash} & N/A & \cellcolor{red!30}-0.079 & N/A & \cellcolor{gray!30}0.05 \\
    \textbf{Claude 3 Haiku} & \cellcolor{yellow!30}$0.444^{*}$ & \cellcolor{gray!30}0.122 & \cellcolor{yellow!30}0.5 & \cellcolor{gray!30}0.4 \\
    \textbf{Claude 4.5 Haiku} & N/A & N/A & N/A & N/A \\
    \textbf{Mistral 8x7B} & N/A & N/A & N/A & N/A \\
    \textbf{Mistral 4 Small} & \cellcolor{yellow!30}$0.448^{**}$ & \cellcolor{red!30}-0.058 & \cellcolor{gray!30}0.417 & \cellcolor{gray!30}0.3 \\
    \textbf{Qwen2.5-7B-Instruct} & \cellcolor{gray!30}$0.383^{**}$ & \cellcolor{gray!30}0.121 & \cellcolor{gray!30}0.375 & \cellcolor{gray!30}0.4 \\
    \textbf{Qwen3-8B} & \cellcolor{yellow!30}$0.523^{***}$& N/A & \cellcolor{gray!30}0.458 & N/A \\ \hline
  \end{tabular}
  \caption{Correlation and Accuracy Between Human and LLM Argument Evaluation. ``Avg.'' correlations are computed with Kendall's tau on the reported average pro - average con message. ``Sign'' looks at the accuracy of the expected sign for each condition. * = p value \textless 0.05, ** = p value \textless 0.01, *** = p value \textless 0.001. For ``Avg.'', red signifies negative relationships, gray signifies weak positive relationships, and yellow signifies moderate positive relationships. For ``Sign'', gray signifies low accuracy.}

  
  \label{tab:humanvllmprovcon}
\end{table*}

We find clear evidence that LLMs deviate from human-like motivated reasoning. Table \ref{tab:humanvllm} columns 2-5 (``Study 1'' - ``Study 4'') show the Kendall's tau between the model and human averages across all 4 studies with regard to their support or change in support. We note no strong positive correlations \cite{Mukaka_correlation} for any model, which we would expect to happen if reasoning patterns matched. This suggests LLMs fail to model human motivated reasoning.




Shifting to analyze change in support when compared to a control, we compute accuracy between the signs of the LLM and human data (Table \ref{tab:accuracysign}), which indicates how often the LLM reflected a similar type of change to humans across conditions. We find low accuracy for both study 1 (except Mistral 8x7B) and study 3. The lack of consistent results across studies highlights the distinction between LLM and human behavior.

Moving to the assessment of argument strength, we find minimal evidence to suggest LLMs replicate human assessments. Study 3 shows significant but moderate correlations (Table \ref{tab:humanvllmprovcon} column 2, ``Study 3 Avg.'') and all models have weak and insignificant correlations with humans in study 4 (Table \ref{tab:humanvllmprovcon} column 3, ``Study 4 Avg.''). This is further supported by examining the accuracy of the expected signs. Across both studies, we find low accuracy across all models, with scores $\leq$ 0.5. (Table \ref{tab:humanvllmprovcon} column 4 and 5, ``Study 3 Sign'' and  ``Study 4 Sign''). Thus, our results suggest that LLMs are not able to assess message strength in ways that resemble humans.


\subsection{LLMs Are Limited in Ability to Assess Relative Argument Strength}
\label{sec:RQ2}



\begin{table}[ht]
\centering
\begin{tabular}
{|l|p{0.1\textwidth}|p{0.1\textwidth}|p{0.1\textwidth}|}
\hline 
 \textbf{Model} & \textbf{Study 3} & \textbf{Study 4} \\
\hline 
\textbf{GPT-4o mini} & \cellcolor{yellow!30} $0.526^{***}$& \cellcolor{gray!30}$0.388^{*}$ \\
\textbf{o3-mini} &\cellcolor{green!30} $0.674^{***}$ & \cellcolor{green!30}$0.66^{***}$ \\
\textbf{Gemini 2.0 Flash} & \cellcolor{gray!30}$0.377^{*}$ & \cellcolor{gray!30}$0.392^{*}$ \\
\textbf{Gemini 2.5 Flash} & N/A & \cellcolor{gray!30}0.192  \\
\textbf{Claude 3 Haiku} & \cellcolor{green!30}$0.649^{***}$ & \cellcolor{gray!30}0.308 \\
\textbf{Claude 4.5 Haiku} & N/A & N/A \\
\textbf{Mistral 8x7B} & N/A & N/A   \\
\textbf{Mistral 4 Small} & \cellcolor{green!30}$0.765^{***}$&  \cellcolor{yellow!30}$0.484^{**}$ \\
\textbf{Qwen2.5-7B-Instruct} & \cellcolor{yellow!30}$0.470^{**}$&  \cellcolor{yellow!30}$0.418^{*} $ \\
\textbf{Qwen3-8B} & \cellcolor{green!30}$0.695^{***}$& N/A \\
\hline \hline
\textbf{Human Baseline} & \cellcolor{green!30}$0.788^{***}$ & \cellcolor{green!30}$0.904^{***}$ \\
\hline 

\end{tabular}
  \caption{Kendall's tau of the Average Opinion in Each Condition and the  Assessment of Argument Strength (Pro - Con) in Each Condition. N/A signifies not enough samples were generated in at least one condition to make a fair comparison. * = p value \textless 0.05, ** = p value \textless 0.01, *** = p value \textless 0.001. Gray signifies weak positive relationships, yellow signifies moderate positive relationships, and green signifies strong positive relationships.}
  \label{tab:spearman_meanvprovcon}
  \vspace{-1em}
\end{table}

To evaluate LLM consistency within answers, we study the relationship between argument assessment and opinion formation for studies that also asked for pro and con argument ratings (studies 3 and 4). We find weak or moderate positive correlations for most models in either one or both studies (Table \ref{tab:spearman_meanvprovcon}). The most robust model is o3-mini, showing strong correlations in both studies. However, all models appear to underperform the human baselines on the same measure, emphasizing how LLMs may provide less consistent evaluations than humans. We also investigate how LLMs assess argument strength directly, discussed further in Appx. Section \ref{sec:argassessments}.

\subsection{Models Selectively ``Opt Out'' of Responding Based on Certain Cues}
\label{sec:RQ3}

The previous results suggest model response opt-out happens quite often for this study replication. Only GPT-4o mini, o3-mini, and Qwen2.5-7B-Instruct provided enough responses for all 4 studies. This suggests some sort of prevention mechanism internally keeping LLMs from providing a response. Notably, models with reasoning capabilities seem to be most prone, with 4 out of 5 failing to respond to at least 2 studies.

We find the opt-out rate is not uniform for most cases where at least 10\% of the samples were missing from at least one condition (Table \ref{tab:chisquare}), considering all conditions independently. To determine if message traits cause increased non-response, we aggregate the missing responses based on manipulation (e.g., Study 1 has 3 ``motivation'' manipulations -- none, directional, and accuracy -- with 15 conditions total. We calculate the total missing responses for each type divided by the number of conditions in that type (here, 5 per motivation) and recompute the Chi-Square measure. This is calculated for all types of manipulations within one study). We find different manipulations affects models in different ways (see Appx. Section \ref{sec:nonresponseanalysis}) with clear patterns emerging based on provided information rather than model family (Table \ref{tab:chisquare_abbrev}). These findings ultimately suggest that while information manipulations may not affect LLM responses, it may affect \textit{which} LLMs respond at all.

\begin{table*}[]
    \centering
    \begin{tabular}{|p{0.1\textwidth}|p{0.3\textwidth}|p{0.5\textwidth}|}
        \hline
        \textbf{Study} & \textbf{Conditions with Significant $\chi^{2}$} & \textbf{Models} \\ \hline
        \textbf{Study 1} & Motivation & \textit{Gemini 2.5 Flash}, Claude 3 Haiku, \textit{Claude 4.5 Haiku} \\ \hline \hline
        \textbf{Study 2} & Information & Gemini 2.0 Flash, \textit{Gemini 2.5 Flash}, \textit{Claude 4.5 Haiku} \\ \hline
        & Technology & Gemini 2.0 Flash, Claude 3 Haiku \\ \hline \hline
        \textbf{Study 3} & Control & Gemini 2.0 Flash, \textit{Gemini 2.5 Flash}, \textit{Claude 4.5 Haiku}, Mistral 8x7B, \textit{Mistral 4 Small}, \textit{Qwen3-8B} \\ \hline
        & Topic & Mistral 8x7B \\ \hline
        & Polarization & Mistral 8x7B \\ \hline
        
    \end{tabular}
    \caption{Conditions with Significant Chi-Square Values for Model Opt-Out5. Italicized models are reasoning models. No categories were significant for any model in Study 4.}
    \label{tab:chisquare_abbrev}
\end{table*}

\subsection{Robustness Testing Via Prompting}

We completed additional testing on a subset of models and studies to measure how outcomes were affected by differences in prompting. In one setting, LLMs were prompted to provide reasoning along with their numeric evaluations across all 4 studies (Appx. Sections \ref{sec:justification}). We additionally recreated a demographically representative sample of one of the only studies which provided original aggregate demographics -- Study 2 -- and provided LLMs with persona inductions (Appx. Section \ref{sec:demographictest}). While LLM behavior differs from the main paper results in both settings, we still find LLMs fail to mimic human motivated reasoning outcomes. This supports the robustness of the main paper findings.

\section{Discussion}

Throughout this study, we explore the impacts of motivated reasoning inductions on LLMs compared to humans and other LLMs.  We find that LLMs do not reason like individuals, highlighting how far LLMs are from accurately representing human reasoning patterns under different motivations (RQ1). Despite having a set of preferences and biases, LLMs do not appear to mimic human-like motivated reasoning (Section \ref{sec:RQ1}). We further find that LLMs struggle to both accurately assess the strength of arguments and incorporate of information into responses (RQ2). While prior work has shown similar pitfalls in LLM argument assessment \cite{Djouvas_Charalampous_Christodoulou_Tsapatsoulis_2024}, we show this is a phenomenon that plagues multiple models across multiple topics (Appx. Section \ref{sec:argassessments}), transcending the effects of motivation. Finally, when models fail to respond is dependent on the information provided to them (Appx. Section \ref{sec:RQ3}), suggesting LLMs can only be used as human proxies under certain settings such as non-controversial topics. While reasoning models tend to opt out more often than non-reasoning models (11 instances compared to 7), we find no relationship between LLM family (Appx. Section \ref{sec:nonresponseanalysis}). Rather, the selective non-response appears to be based on the conditions of the studies, as shown with clear model trends in Studies 1, 3, and 4 (Table \ref{tab:chisquare}).

The finding that base LLMs do not inherently pursue motivated reasoning from inductions deviates from prior works which show LLMs pursue different motivations when given personalities \cite{Dash_Reymond_Spiro_Caliskan_2025, Cho_Hoyle_Hermstrwer_2025}. By simplifying the experiments and focusing on base LLMs, our work suggests that any ``motivations'' LLMs embody may be heavily dependent on the personas. This emphasizes the importance of persona selection and group representation in these settings. However, we still find LLMs do not replicate human motivated reasoning even when given various demographics to consider (Section \ref{sec:demographictest}), suggesting they are not affected by motivational cues even when given personas. Rather than prescribing a reason for the difference in these behaviors (such as sycophancy \cite{Pitre_Ramakrishnan_Wang_2025, Fanous_Goldberg_Agarwal_Lin_Zhou_Xu_Bikia_Daneshjou_Koyejo_2025}, prior information, or black-box system directives), we highlight the gap between base LLM behavior, LLM with persona behavior, and human behavior within the motivated reasoning space.

The findings of this work can be used to further advance social science and computer science studies as both fields shift towards supplementing or conducting surveys using LLMs. Collecting representative human samples is often costly \cite{NSF21-601}, and the same data using LLMs can be generated more quickly and cheaply. However, results from human simulation of these surveys is somewhat mixed \cite{Zhang_Xu_Alvero_2025, Argyle_Busby_Fulda_Gubler_Rytting_Wingate_2023}, covering less diversity when compared to human responses \cite{Bisbee_Clinton_Dorff_Kenkel_Larson_2024, Zhang_Xu_Alvero_2025}, even when given personas to mimic. Motivations may be a factor that drives certain respondents' responses when they enter these surveys, shaping the diversity of the sample population. While prompting LLMs with motivation cues could have potentially solved this issue, we find this is not the case, suggesting LLMs may not be accurate forecasters of public opinion when motivated reasoning is involved. Additionally, prompting for feedback on more divisive topics may lead LLMs to abstain from responding, making it harder to utilize them as forecasters in controversial settings.


This work also has implications for the AI and NLP communities who utilize LLMs for automating certain tasks, such as argument assessment. Given the limitations of LLMs in assessing argument strength under different conditions (Appx. Section \ref{sec:argassessments}), researchers should consider building advanced LLM systems, utilizing techniques such as few-shot prompting or reasoning, in order to improve reliability of such systems. Additionally, incorporating a mixture of human and LLM assessments could lead to more grounded systems with a higher number of samples. The incorporation of guardrails and internal checks will help create more robust and more reliable systems for these tasks. 

Considering our findings, future work could explore manipulating personality and motivational inductions for LLM survey tasks to better emulate human-like reasoning beyond the checks presented in this work. Furthermore, researchers could examine different techniques of boosting argument assessment performance to improve accuracy. Exploring these topics in the future could potentially align LLMs with human motivation and preferences for survey and annotation tasks.

\section*{Limitations}

\textbf{LLM-Based Limitations.} First, we test a small subset of possible parameters and models. While we intend to offer a comprehensive view of LLM political motivated reasoning, we acknowledge this work could cover more models under different parameter specifications. Second, we test variations in prompting on subsets of the models and studies. Future work could explore variations in prompt directive on a larger set of studies. While we test a subset of studies and models with demographics, this could be expanded. Additionally, the reasoning provided is not studied, and future work could analyze the text-based reasoning to see how reasoning patterns differ when different numeric answers are provided. 


\noindent
\textbf{Study Replication-Based Limitations.}   We selected a subset of motivated reasoning work to study. Motivated reasoning can occur in fields outside of politics such as health \cite{Kunda_1990}. This work could be expanded by looking at studies across a variety of domains or expanding the amount of studies considered within the domain of political science. Furthermore, the studies replicated in this work were published in 2013 - 2016. It is possible that at least some of the models will have some information in their training around some of these policies. Future work could explore how LLMs respond in similar motivated reasoning scenarios but on more recent events, particularly those that occur outside of its knowledge cutoff. In terms of replicating the studies, we opted to not recreate the full sample size but to downsample based on the smallest condition from all studies. While this was done to minimize cost and resource constraints, given we comparing averages, preserving the exact sample size may not be necessary for most analyses. Future work could replicate the studies in full to see how a larger sample size shapes results.


\bibliography{llmmotivatedreasoning}

\newpage

\appendix

\section{LLM Parameters and Prompts}
\label{sec:llmparam}

\subsection{Prompts Used Within the Study}
Because all text prompts regarding (1) information provided to participants about the topic, (2) motivations inductions, and (3) questions asked to participants are easily accessible in the original manuscripts and their appendices, they will not be provided here directly. For Study 1 \cite{Bolsen_Druckman_Cook_2014}, all information is provided in the main paper. In study 2 \cite{Bolsen_Druckman_2015}, all message information is provided in the appendix. In study 3 \cite{Druckman_Peterson_Slothuus_2013}, all information is provided in the main paper. In study 4 \cite{Mullinix_2016}, all information is provided in the supplementary materials. For a high-level summary of manipulations per study, see Table \ref{tab:studyoverview}.

\subsection{API-Based Models}
Of the models considered in this study, GPT 4o-mini, o3-mini, Gemini 2.0 Flash, Gemini 2.5 Flash, Claude 3 Haiku, Claude 4.5 Haiku, Mistral 8x7B, and Mistral 4 Small were called using the publicly available API each company offers. For each non-reasoning model, the ``max\_tokens'' was set to 1000, while reasoning models were given 2000 tokens. Reasoning models were given additional tokens to account for the behind-the-scenes reasoning before giving a response. Preliminary tests showed reasoning models struggled to respond at high rates with only 1000 tokens. For the OpenAI family of models, the ``temperature'' and ``top\_p'' were set to 1, following their default setting at the time of generation\footnote{\url{https://platform.openai.com/docs/api-reference/chat}}. For Gemini models, ``temperature'' was set to 1, ``top\_p'' was set to 0.95, and ``top\_k'' was set to 64, following their default setting at the time of generation\footnote{\url{https://docs.cloud.google.com/vertex-ai/generative-ai/docs/models/gemini/2-0-flash}}. For the Anthropic family of models, the ``temperature'' was set to 1, following their default setting at the time of generation\footnote{\url{https://platform.claude.com/docs/en/api/beta/messages/create}}. For Mistral, ``frequency\_penalty'' was set to 0, ``presence\_penalty'' was set to 0, and ``top\_p'' was set to 1 (opting to set this over ``temperature''), following their default setting at the time of generation\footnote{\url{https://docs.mistral.ai/api}}. Mistral 4 Small's ``reasoning\_effort'' was set to ``high'' to activate reasoning capabilities. The specific version of the model is listed below:

\begin{itemize}
    \item gpt-4o-mini-2024-07-18
    \item o3-mini-2025-01-31
    \item gemini-2.0-flash-001
    \item gemini-2.5-flash
    \item claude-3-haiku-20240307
    \item claude-haiku-4-5-20251001
    \item open-mixtral-8x7b
    \item mistral-small-2603
\end{itemize}

\subsection{On-Device Models}
Qwen2.5-7B-Instruct and Qwen3-8B were run by directly accessing the model rather than through API provided by the parent company. For the main results and the ``reasoning ablation'', Qwen2.5-7B-Instruct was run by adapting the code provided on Huggingface\footnote{\url{https://huggingface.co/Qwen/Qwen2.5-7B-Instruct}} and computed using 2 NVIDIA RTX A6000s. For the demographic ablation, Qwen2.5-7B-Instruct was run through ollama based API on the previously described server. All Qwen3-8B results were run through ollama on the same server.

\subsection{Dropped Models}
In addition to the 5 models analyzed in this study, 5 other models were considered. OpenAI's o1-mini, while used previously to benchmark political bias \cite{Yang_Menczer_2025} was depreciated\footnote{\url{https://platform.openai.com/docs/deprecations}} and full data was not collected. Additionally, Anthropic's claude3.5haiku was also depreciated during the course of the study\footnote{\url{https://platform.claude.com/docs/en/about-claude/model-deprecations}}. Similarly, Gemini 1.5 Flash, also used in assessing political bias \cite{Proma_Pate_Druckman_Ghoshal_He_Hoque_2025} was also depreciated\footnote{\url{https://ai.google.dev/gemini-api/docs/changelog}}. Google's Gemini 2.0 Flash-Exp, an experimental ``thinking'' version of Gemini 2.0 Flash, has a response per day maximum of 500 for lower tiers, making it unable to complete one full study in a day. Deepseek models were also considered but ultimately dropped as well.

\section{Additional Details on Study Replication}
\label{sec:appxc}

\subsection{Details of Each Condition Within Each Study}
\label{sec:studydetails}

Table \ref{tab:studyoverview} discusses the topics, outcomes, motivation and additional manipulations, and result presentation of the original studies. The below sections provide more detail regarding motivation manipulations.

\subsubsection{The Influence of Partisan Motivated Reasoning on Public Opinion (Bolsen, Druckman, and Cook; 2014)}

This study manipulates motivation in 3 ways: an ``accuracy'' prompt, a ``directional'' prompt, or no motivational prompt. It also manipulates party endorsement in 5 ways: ``same party'' (as the particular individual in the study), ``different party'' (as the particular individual in the study), ``consensus'' or strong support from both parties, ``cross-partisan'' or weak support from both parties, or none. This results in 15 pair-wise combinations. The ``no motivation'' prompt with the ``accuracy'' motivation acts as the control for this study. See Table \ref{tab:1.1samples} for a list of all conditions.

\subsubsection{Counteracting the Politicization of Science (Bolsen and Druckman; 2015)}
The study presenting scientific information, polarized scientific information (sewing doubt in scientific information), and strategies to mitigate motivated reasoning when encountering polarized scientific information (such as warnings and corrections) on the assessment of information in two domains: fracking and carbon nanotubes (CNTs). Here, politicizing the issue is meant to invoke directional motivated reasoning. There are 12 conditions total, with ``no information'' for both ``fracking'' and ``CNTs'' acting as controls. See Table \ref{tab:1.2samples} for a list of all conditions.

\subsubsection{How Elite Partisan Polarization Affects Public Opinion Formation (Druckman, Peterson, Slothuus; 2013)}
This study manipulates argument strength and polarization across conditions. To manipulate argument strength, they provide either strong or weak arguments attached to either pro or con arguments, resulting in 4 different combinations (Pro Strong Con Strong, Pro Strong Con Weak, Pro Weak Con Strong, Pro Weak Con Weak). They additionally manipulate party polarization cues, offering no party cues, non-polarized party cues, and polarized party cues. This occurs for both the drilling and DREAM conditions. This study has 26 conditions, where ``no information'' and ``no party endorsements'' acting as controls. See Table \ref{tab:1.3samples} for a list of all conditions.

\subsubsection{Partisanship and Preference Formation: Competing Motivations, Elite Polarization, and Issue Importance (Mullinix; 2016)}
The study manipulates polarization, endorsement cues, and personal importance of the outcome. Polarization can be low or high, leading to heightened attention to partisanship. Importance can be low or high, making the impact of the policies more salient to the individual. Endorsement can be none, traditional (based on common knowledge of party stances) or reversed. There are 20 conditions total, with two controls:  ``Low polarization, low issue importance, no party endorsement'' and ``No polarization cue, no issue importance cue, no party endorsement'' for both taxes and the SSA. See Table \ref{tab:1.4samples} for a list of all conditions.

\subsection{Number of Samples}

The number of samples from the original studies along with the attempted amount of generations are reported in Tables \ref{tab:1.1samples}, \ref{tab:1.2samples}, \ref{tab:1.3samples}, and \ref{tab:1.4samples}. The proportional downsampling preserves study structure by keeping conditions weighted relatively the same while requiring less cost and resources to generate the results. These results are averaged and compared against human averages, and so preserving the exact sample size may not be necessary to conduct this analysis.

However, the total amount of usable responses was not always equal to this number. For example, in certain conditions, certain models would be more likely to resist answering the questions, citing being an AI model that does not have opinions on certain issues. In study 1, Claude 3.5 Haiku only answered 19 out of 48 of the attempted generations for the ``Cross-partisan endorsement, no motivation'' condition. In study 3, GPT-4o mini only answered 29 of the attempted 47 ``Control, DREAM'' condition and Claude 3.5-haiku only answered 6 of the ``Control, Drilling'' and none of the ``Control, DREAM'' condition. These were the only large instances of data dropping, and why Claude 3.5-haiku was removed from study 1 and Claude 3.5-haiku and GPT 4o-mini were dropped from the support relative to the control analysis for study 3.

\label{sec:numsample}

\begin{table*}
  \centering
  \begin{tabular}{|p{0.1\linewidth}|p{0.1\linewidth}|p{0.1\linewidth}|p{0.15\linewidth}|p{0.15\linewidth}|p{0.1\linewidth}|p{0.2\linewidth}|} \hline
    \textbf{Study} & \textbf{Topic} & \textbf{Outcomes} & \textbf{Directional Manipulations} & \textbf{Accuracy Manipulations} & \textbf{Additional Manipulations} & \textbf{Result Presentation} \\ \hline
    \textbf{Study 1} & 2007 Energy Independence Act & Opinion (1-7) & Party identification justification & Reason justification & N/A & Percent change compared to control \\ \hline
    \textbf{Study 2} & Fracking or carbon nanotubes (CNTs) & Opinion (1-7) & Politicizing scientific information & Adding warnings, corrections, and accuracy prompts to politicized scientific information & N/A & Percent responses of 5 or greater (support) \\ \hline
    \textbf{Study 3} & Drilling or DREAM Act & Opinion (1-7), Argument strength (1-7) & Polarization of an issue & No polarization & Strength of pro and con arguments provided & Percent change compared to control (opinion), Difference in pro and con argument ratings (argument assessment) \\ \hline
    \textbf{Study 4} & Tax cut or SSA & Opinion (0-10), Argument strength (1-7) & Polarization of an issue, Issue importance & No polarization, no issue importance, and no endorsements given & Party endorsement & Mean assessment (opinion), Difference in pro and con argument ratings (argument assessment) \\ \hline
\end{tabular}
\caption{High-Level Breakdown of Study Variables.}
\label{tab:studyoverview}
\end{table*}

\begin{table*}
  \centering
  \begin{tabular}{|p{0.6\linewidth}|p{0.15\linewidth}|p{0.15\linewidth}|p{0.1\linewidth}|}
    \hline
\textbf{Condition} & \textbf{Original N} & \textbf{Generations} & \textbf{Figure Key}\\
    \hline
No endorsement, no motivation &  73 & 53 & A\\    \hline
     No endorsement, directional motivation & 77 & 56 & B
    \\    \hline
      No endorsement, accuracy motivation (Control) & 65 & 47 & C
    \\    \hline
     Same party, no motivation & 73 & 53 & D
    \\    \hline
     Same party, directional motivation & 79 & 57 & E
    \\    \hline
     Same party, accuracy motivation & 62 & 45 & F
    \\    \hline
     Different party, no motivation & 80 & 58 & G
    \\    \hline
     Different party, directional motivation &  76 & 55 & H
    \\    \hline
     Different party, accuracy motivation & 76 & 55 & I
    \\    \hline
     Consensus endorsement, no motivation & 64 & 46 & J
    \\    \hline
     Consensus endorsement, directional motivation & 77 & 56 & K
    \\    \hline
     Consensus endorsement, accuracy motivation & 73 & 53 & L
    \\    \hline
     Cross-partisan endorsement, no motivation & 66 & 48 & M 
    \\    \hline
     Cross-partisan endorsement, directional motivation & 68  & 49 & N
    \\     \hline
     Cross-partisan endorsement, accuracy motivation & 61 & 44 & O          \\\hline
  \end{tabular}

  \caption{``The Influence of Partisan Motivated Reasoning'' Original Samples and Downsampled Samples. Includes key for figure reference.}
  \label{tab:1.1samples}
\end{table*}

\begin{table*}
  \centering
  \begin{tabular}{|p{0.6\linewidth}|p{0.15\linewidth}|p{0.15\linewidth}|p{0.1\linewidth}|}
    \hline
\textbf{Condition} & \textbf{Original N} & \textbf{Generations} & \textbf{Figure Key}\\
    \hline
No information, fracking &  186 &  44 & A \\    \hline
Scientific information, fracking & 189 &  45 & B
    \\    \hline
Politicized scientific information, fracking & 190 &  45 & C
    \\    \hline
Warning + politicized scientific information, fracking & 185 & 44 & D
    \\    \hline
Politicized scientific information + correction, fracking & 213 &  51 & E
    \\    \hline
Accuracy motivation + politicized scientific information + correction, fracking & 189 &  45 & F
    \\    \hline
No information, CNT & 206 &  49 & G
    \\    \hline
Scientific information, CNT &  194 &  46 & H
    \\    \hline
Politicized scientific information, CNT & 192 & 46 & I
    \\    \hline
Warning + politicized scientific information, CNT & 196 & 47 & J
    \\    \hline
Politicized scientific information + correction, CNT & 203 &  48 & K
    \\    \hline
Accuracy motivation + politicized scientific information + correction, CNT & 196 & 47 & L \\
    \hline
  \end{tabular}

  \caption{``Counteracting the Politicization of Science'' Original Samples and Downsampled Samples. Includes key for figure reference.}
  \label{tab:1.2samples}
\end{table*}

\begin{table*}
  \centering
  \begin{tabular}{|p{0.6\linewidth}|p{0.15\linewidth}|p{0.15\linewidth}|p{0.1\linewidth}|}
    \hline
\textbf{Condition} & \textbf{Original N} & \textbf{Generations} & \textbf{Figure Key}\\
    \hline
Control, drilling &  47 &  47 & A \\    \hline
Pro Strong Con Strong No Parties, Drilling & 48 & 48 & B
    \\    \hline
Pro Strong Con Weak No Parties, Drilling & 55 & 55 & C
    \\    \hline
Pro Weak Con Strong No Parties, Drilling & 51 & 51 & D
    \\    \hline
Pro Weak Con Weak No Parties, Drilling & 46 &  46 & E
    \\    \hline
Pro Strong Con Strong Non Polarized Parties, Drilling & 51 &  51 & F
    \\    \hline
Pro Strong Con Weak Non Polarized Parties, Drilling & 49 &  49 & G
    \\    \hline
Pro Weak Con Strong Non Polarized Parties, Drilling &  50 &  50 & H
    \\    \hline
Pro Weak Con Weak Non Polarized Parties, Drilling & 51 & 51 & I
    \\    \hline
Pro Strong Con Strong Polarized Parties, Drilling & 54 & 54 & J
    \\    \hline
Pro Strong Con Weak Polarized Parties, Drilling & 49 &  49 & K
    \\    \hline
Pro Weak Con Strong Polarized Parties, Drilling & 45 & 45 & L \\
    \hline
Pro Weak Con Weak Polarized Parties, Drilling & 50 & 50 & M \\
    \hline
Control, DREAM & 47 & 47 & N \\
    \hline
Pro Strong Con Strong No Parties, DREAM & 46 & 46 & O \\
    \hline
Pro Strong Con Weak No Parties, DREAM & 55 & 55 & P\\
    \hline
Pro Weak Con Strong No Parties, DREAM & 49 & 49 & Q\\
    \hline
Pro Weak Con Weak No Parties, DREAM & 50 & 50 & R\\
    \hline
Pro Strong Con Strong Non Polarized Parties, DREAM & 51 & 51 & S \\
    \hline
Pro Strong Con Weak Non Polarized Parties, DREAM & 44 & 44 & T\\
    \hline
Pro Weak Con Strong Non Polarized Parties, DREAM & 56 & 56 & U\\
    \hline
Pro Weak Con Weak Non Polarized Parties, DREAM & 50 & 50 &  V \\
    \hline
Pro Strong Con Strong Polarized Parties, DREAM & 52 & 52 & W \\
    \hline
Pro Strong Con Weak Polarized Parties, DREAM & 52 & 52 & X\\
    \hline
Pro Weak Con Strong Polarized Parties, DREAM & 50 & 50 & Y \\
    \hline
Pro Weak Con Weak Polarized Parties, DREAM & 44 & 44 & Z\\
    \hline
  \end{tabular}

  \caption{``How Elite Partisan Polarization Affects Public Opinion Formation'' Original Samples and Downsampled Samples. Includes key for figure reference.}
  \label{tab:1.3samples}
\end{table*}

\begin{table*}
  \centering
  \begin{tabular}{|p{0.6\linewidth}|p{0.125\linewidth}|p{0.125\linewidth}|p{0.1\linewidth}|}
    \hline
\textbf{Condition} & \textbf{Original N} & \textbf{Generations} & \textbf{Figure Key}\\
    \hline
Low polarization, low importance, no endorsement, tax (Control 1) &  194 & 48 & A\\    \hline
No polarization, no importance, no endorsement, tax (Control 2) & 191 & 47 & B
    \\    \hline
Low polarization, low importance, traditional endorsement, tax & 190 & 47 &C
    \\    \hline
Low polarization, high importance, traditional endorsement, tax & 209 & 51 &D
    \\    \hline
High polarization, low importance, traditional endorsement, tax & 184 &  45 & E
    \\    \hline
High polarization, high importance, traditional endorsement, tax & 214 &  53 &F
    \\    \hline
Low polarization, low importance, reversed endorsement, tax & 186 &  46 &G 
    \\    \hline
Low polarization, high importance, reversed endorsement, tax &  193 & 47 & H
    \\    \hline
High polarization, low importance, reversed endorsement, tax & 196 & 48 & I
    \\    \hline
High polarization, high importance, reversed endorsement, tax & 197 &48 & J
    \\    \hline
Low polarization, low importance, no endorsement, education (Control 1) & 189 &  46 & K
    \\    \hline
No polarization, no importance, no endorsement, education (Control 2) & 190 & 47 & L\\
    \hline
Low polarization, low importance, traditional endorsement, education & 184 & 45 & M\\
    \hline
Low polarization, high importance, traditional endorsement, education & 203 & 50 & N\\
    \hline
High polarization, low importance, traditional endorsement, education & 182 & 45 & O \\
    \hline
High polarization, high importance, traditional endorsement, education & 206 &  53 & P\\
    \hline
Low polarization, low importance, reversed endorsement, education & 179 & 44 & Q\\
    \hline
Low polarization, high importance, reversed endorsement, education & 189 & 46 & R \\
    \hline
High polarization, low importance, reversed endorsement, education & 190 & 47 & S\\
    \hline
High polarization, high importance, reversed endorsement, education & 191 & 47 & T\\

    \hline
  \end{tabular}

  \caption{``Partisanship and preference formation: Competing motivations, elite polarization, and issue importance'' Original Samples and Downsampled Samples.}
  \label{tab:1.4samples}
\end{table*}

\section{Additional Results}
\label{sec:addres}

\subsection{Main Paper Supporting Figures and Tables}
\label{sec:mainpaper}

\begin{figure*}
    \centering

    \includegraphics[width=0.9\linewidth]{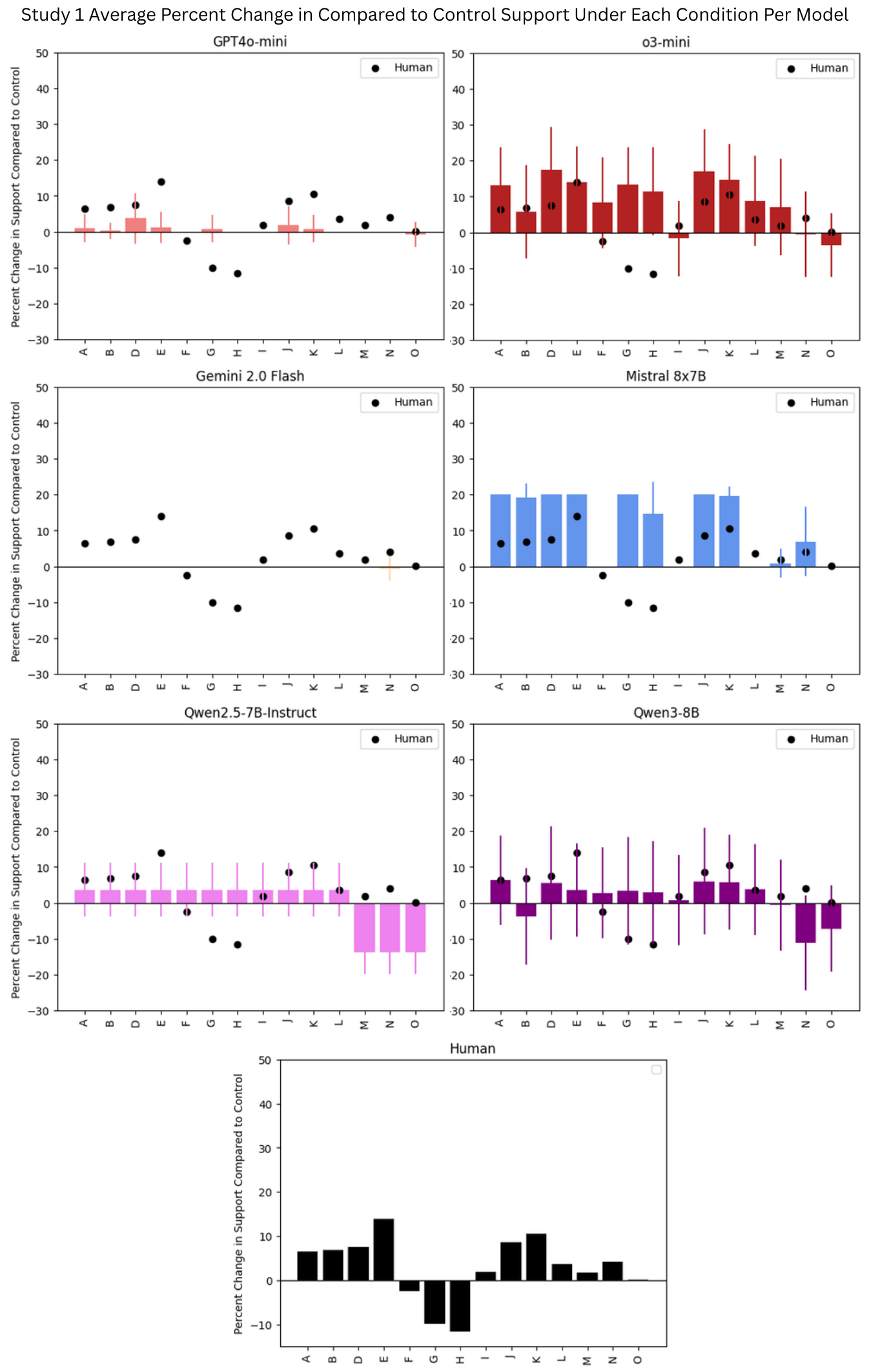}
    \caption{Results for the average percent change in support compared to the control, partitioned by model and condition, compared against the reported human percent change in support (Study 1). Human results from the original paper provided in black for reader convenience.}
    \label{fig:1.1_percentchangepercondition.pdf}
\end{figure*}

\begin{figure*}
    \centering
    \includegraphics[width=0.95\linewidth]{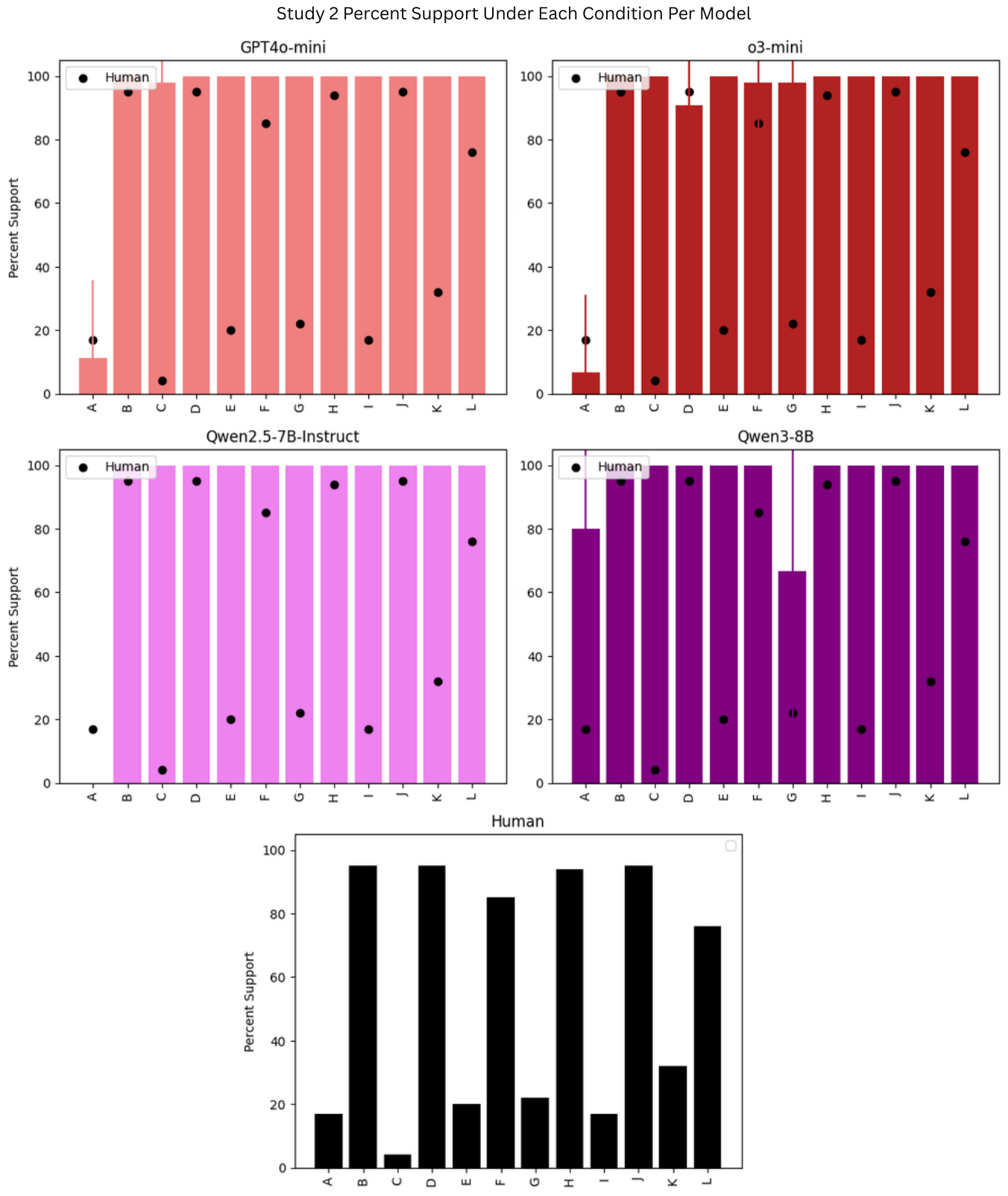}
    \caption{Results for the average percent of support, partitioned by model and condition, compared against the reported human percent change in support (Study 2). Human results from the original paper provided in black for reader convenience.}
    \label{fig:1.2_percentchangepercondition.pdf}
\end{figure*}

\begin{figure*}
    \centering
    \includegraphics[width=0.95\linewidth]{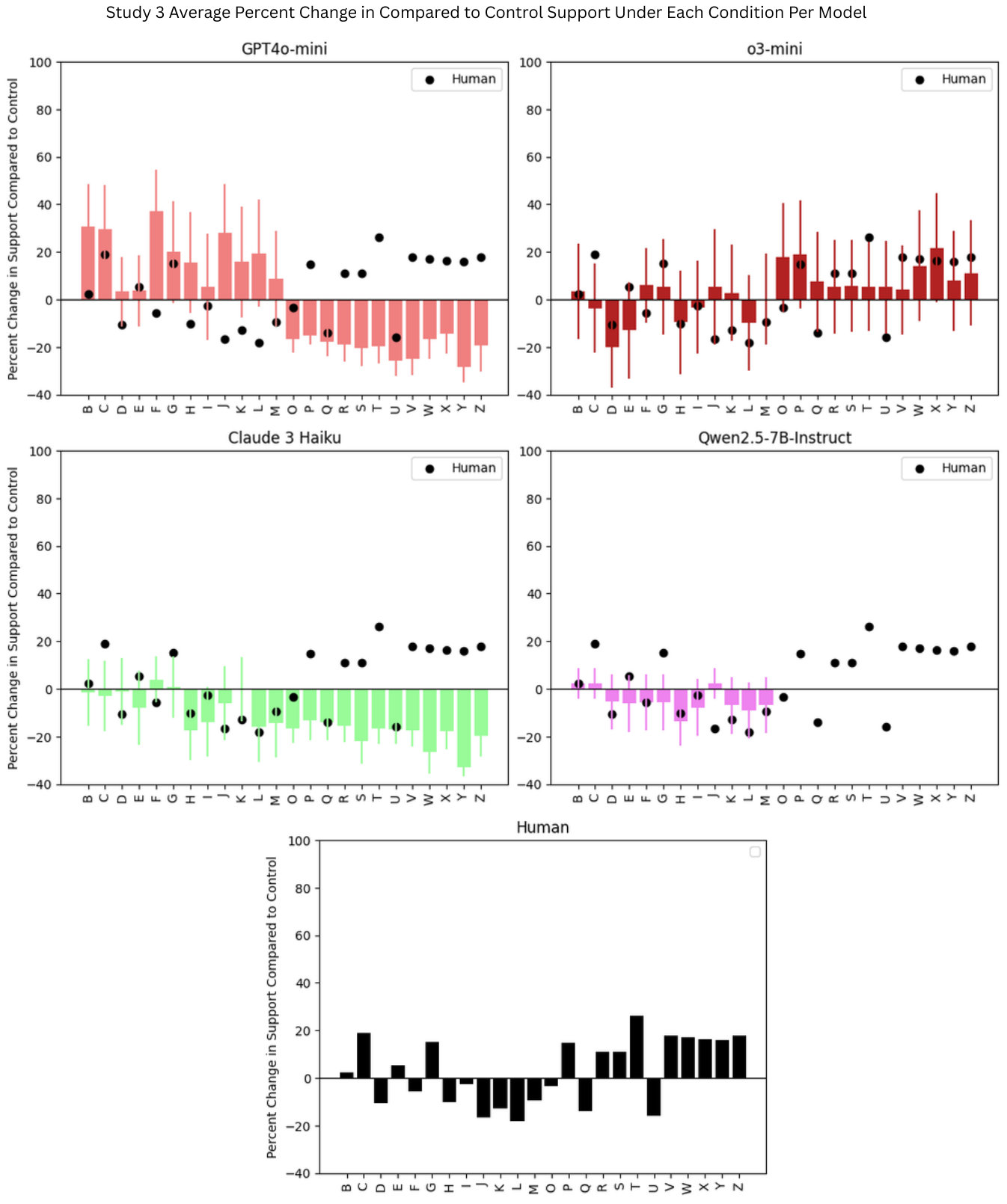}
    \caption{Results for the average percent change in support compared to the control, partitioned by model and condition, compared against the reported human percent change in support (Study 3). Human results from the original paper provided in black for reader convenience.}
    \label{fig:1.3_percentchangepercondition.pdf}
\end{figure*}

\begin{figure*}
    \centering
    \includegraphics[width=0.87\linewidth]{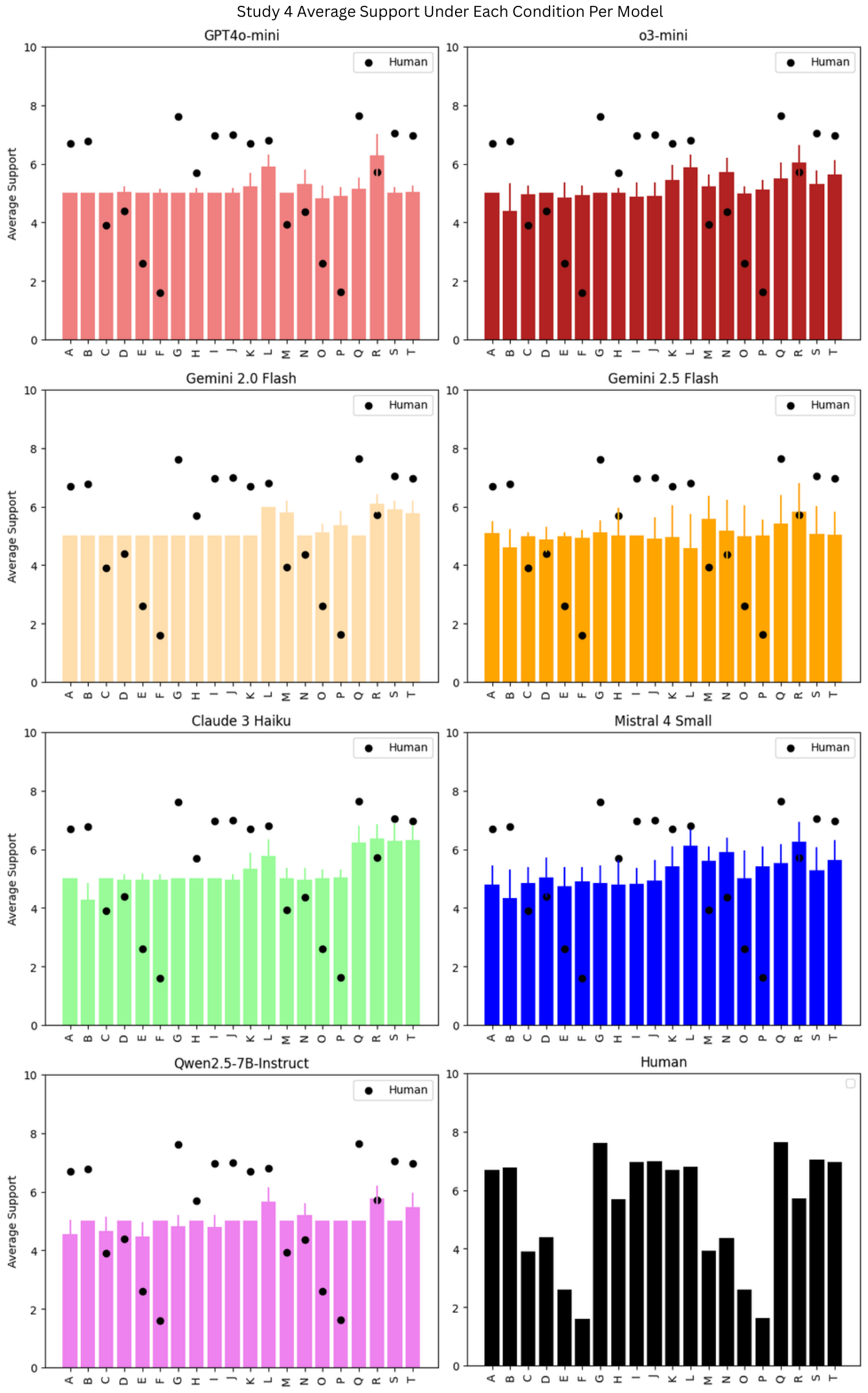}
    \caption{Results for average support, partitioned by model and condition, compared against the reported human percent change in support (Study 4). Human results from the original paper provided in black for reader convenience.}
    \label{fig:1.4_averagesupportpercondition.pdf}
\end{figure*}

\begin{figure*}
    \centering
    \includegraphics[width=0.8\linewidth]{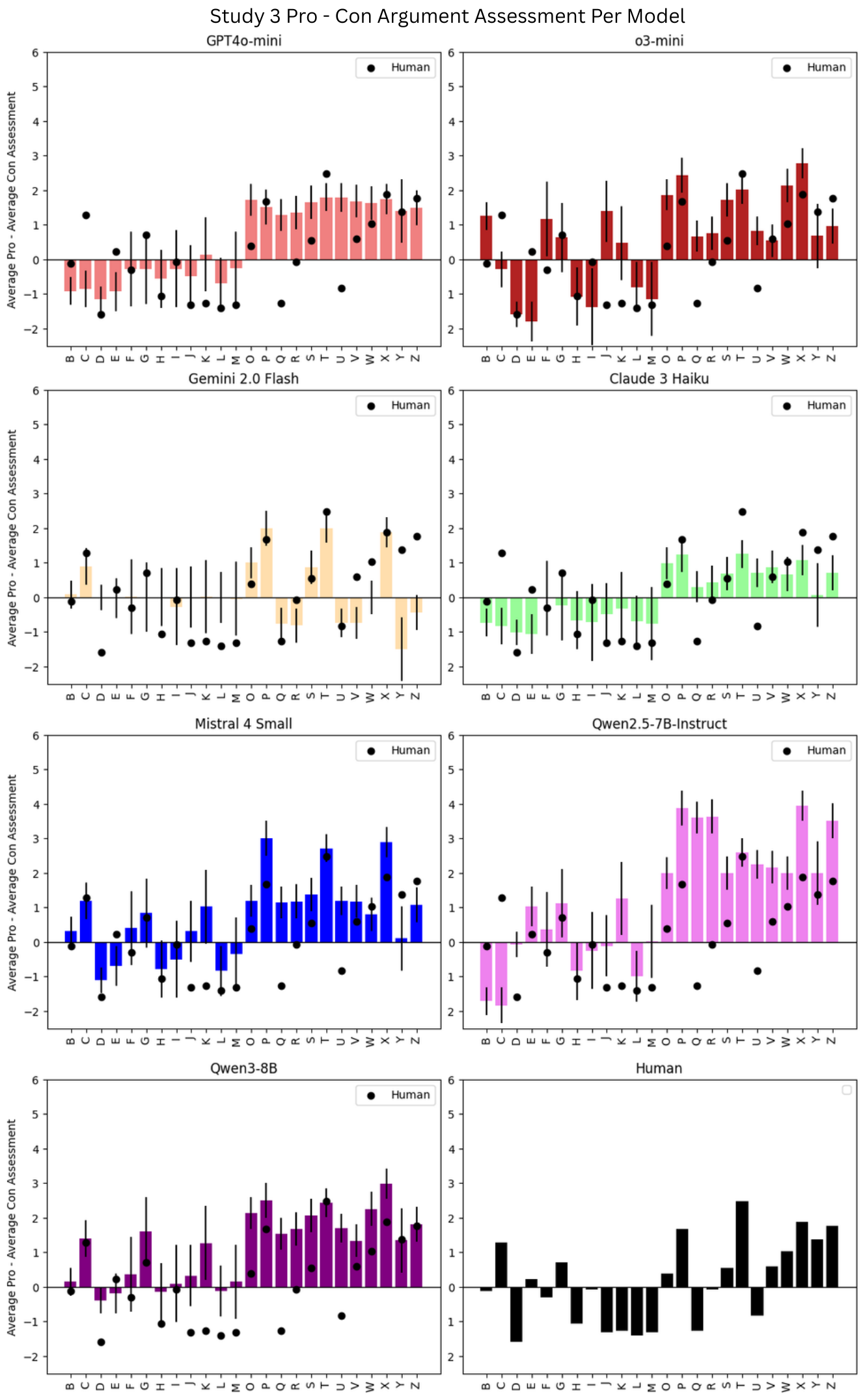}
    \caption{Results for average assessment of pro arguments - average assessment of con arguments, partitioned by model and condition, compared against the reported human percent change in support (Study 3). Human results from the original paper provided in black for reader convenience.}
    \label{fig:1.3_provcon_permodel.pdf}
\end{figure*}

\begin{figure*}
    \centering
    \includegraphics[width=0.95\linewidth]{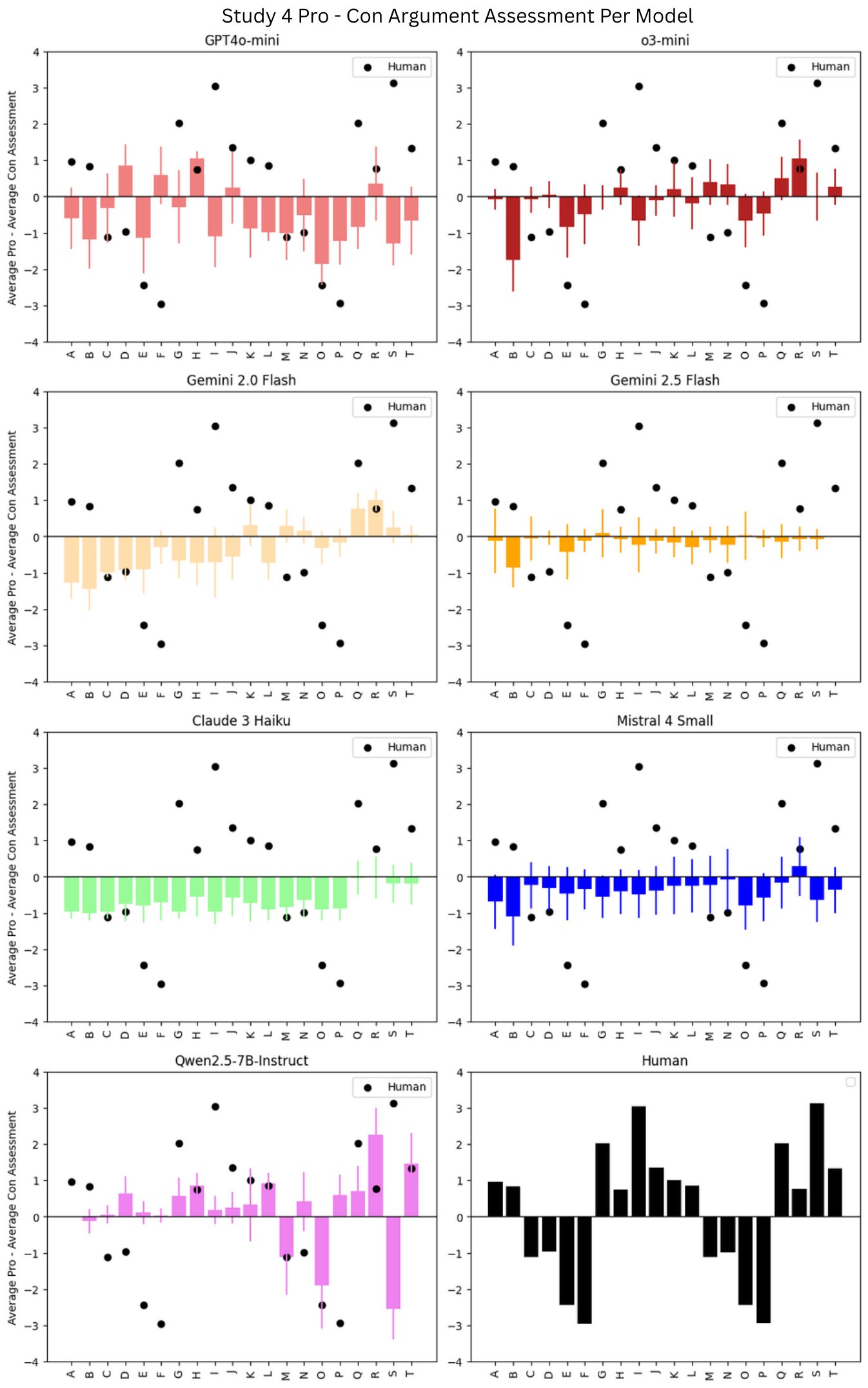}
    \caption{Results for average assessment of pro arguments - average assessment of con arguments, partitioned by model and condition, compared against the reported human percent change in support (Study 4). Human results from the original paper computed using the publicly available replication data \cite{Mullinix_2018}.}
    \label{fig:1.4_provcon_permodel.pdf}
\end{figure*}

Figures \ref{fig:1.1_percentchangepercondition.pdf}, \ref{fig:1.2_percentchangepercondition.pdf}, \ref{fig:1.3_percentchangepercondition.pdf}, and \ref{fig:1.4_averagesupportpercondition.pdf} show the results of mean support responses for each condition. Additionally, the reported human baseline is also provided. Figures \ref{fig:1.3_provcon_permodel.pdf} and \ref{fig:1.4_provcon_permodel.pdf} show the results of the pro - con means for studies 3 and 4, along with the human baseline.

\begin{figure*}
    \centering
    \includegraphics[width=0.95\linewidth]{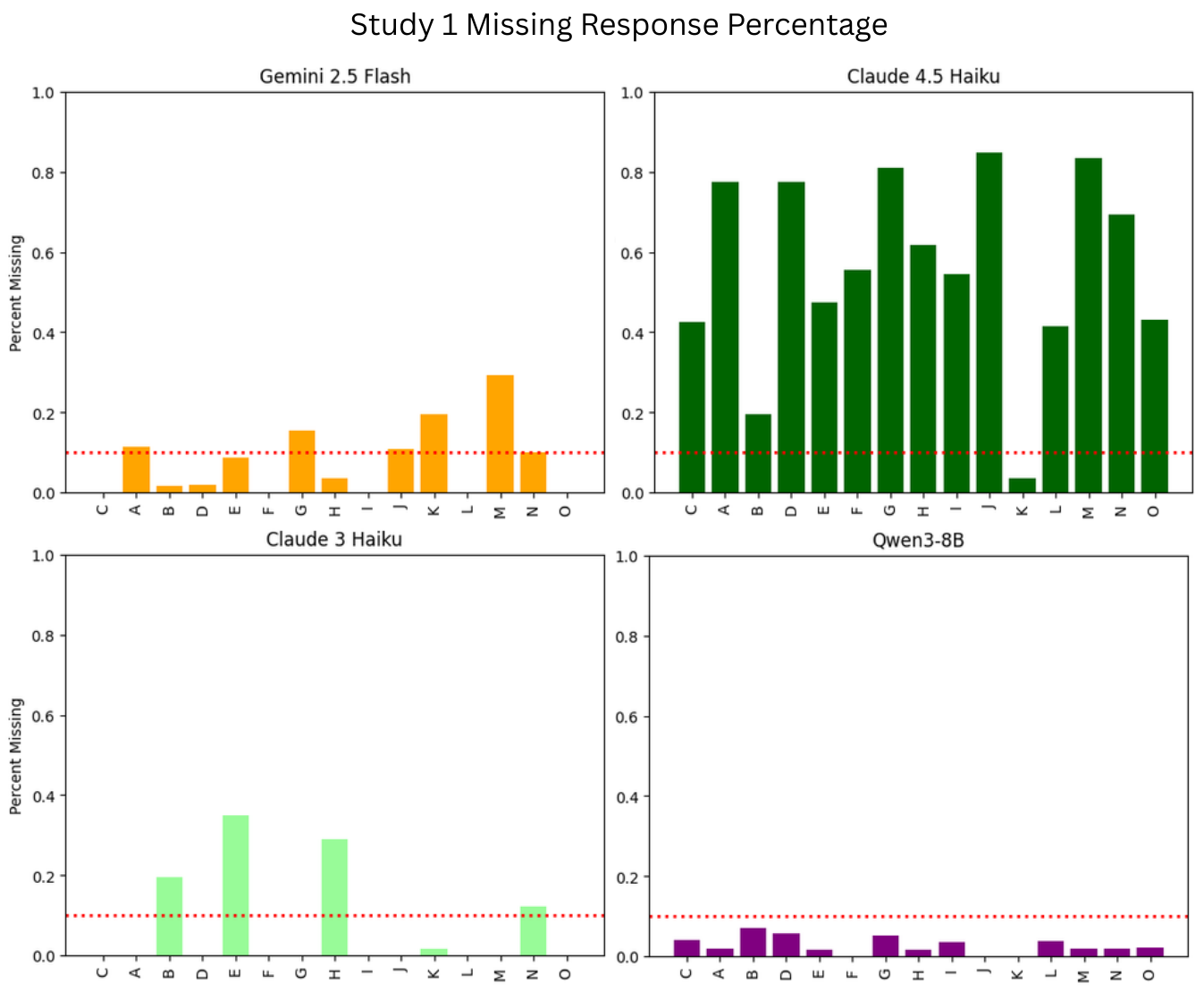}
    \caption{Percentage of missing data partitioned by model and condition in Study 1. 10\% denoted by red line.}
    \label{fig:study1_missing.pdf}
\end{figure*}

\begin{figure*}
    \centering
    \includegraphics[width=0.95\linewidth]{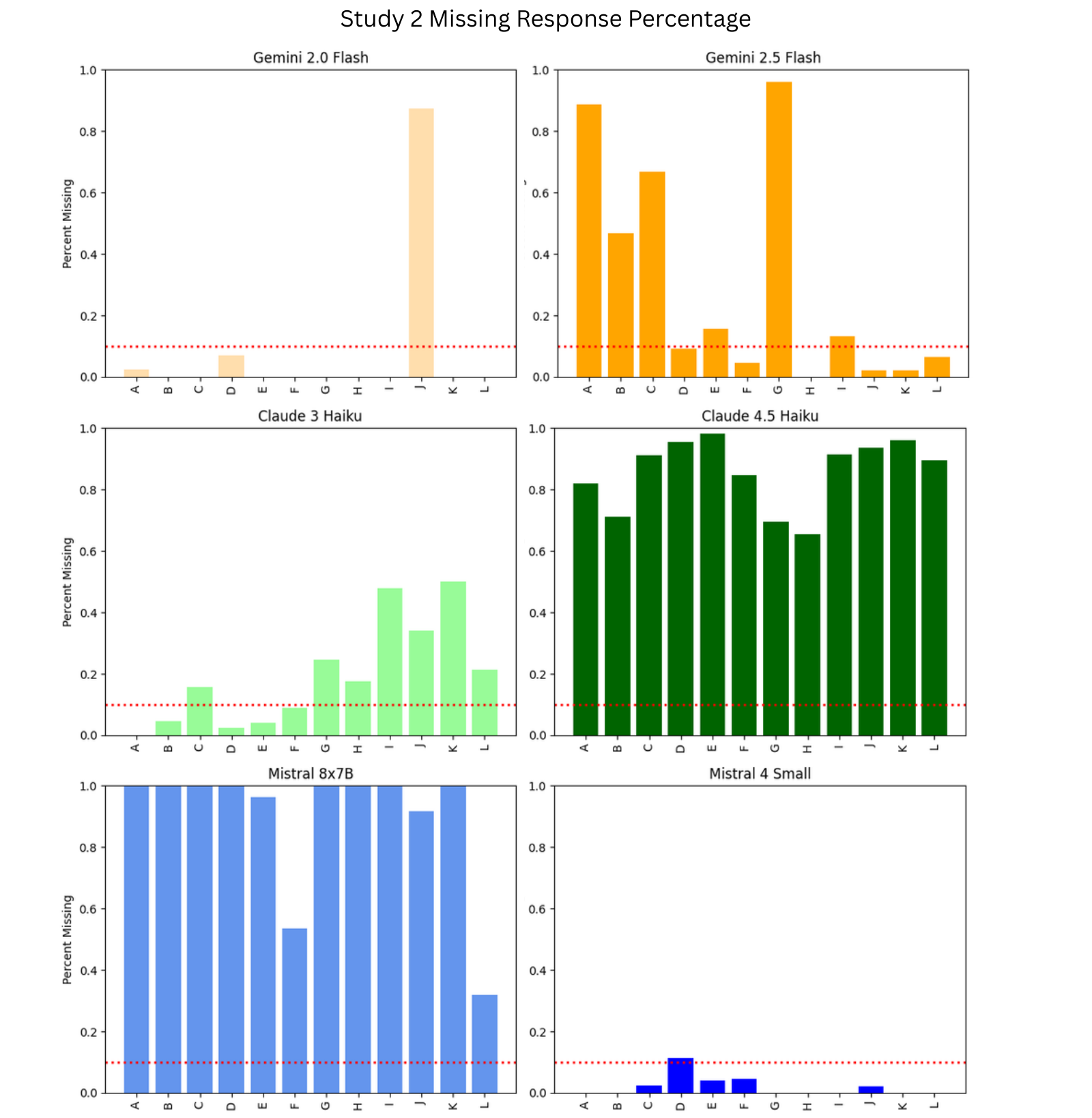}
    \caption{Percentage of missing data partitioned by model and condition in Study 2. 10\% denoted by red line.}
    \label{fig:study2_missing.pdf}
\end{figure*}

\begin{figure*}
    \centering
    \includegraphics[width=0.95\linewidth]{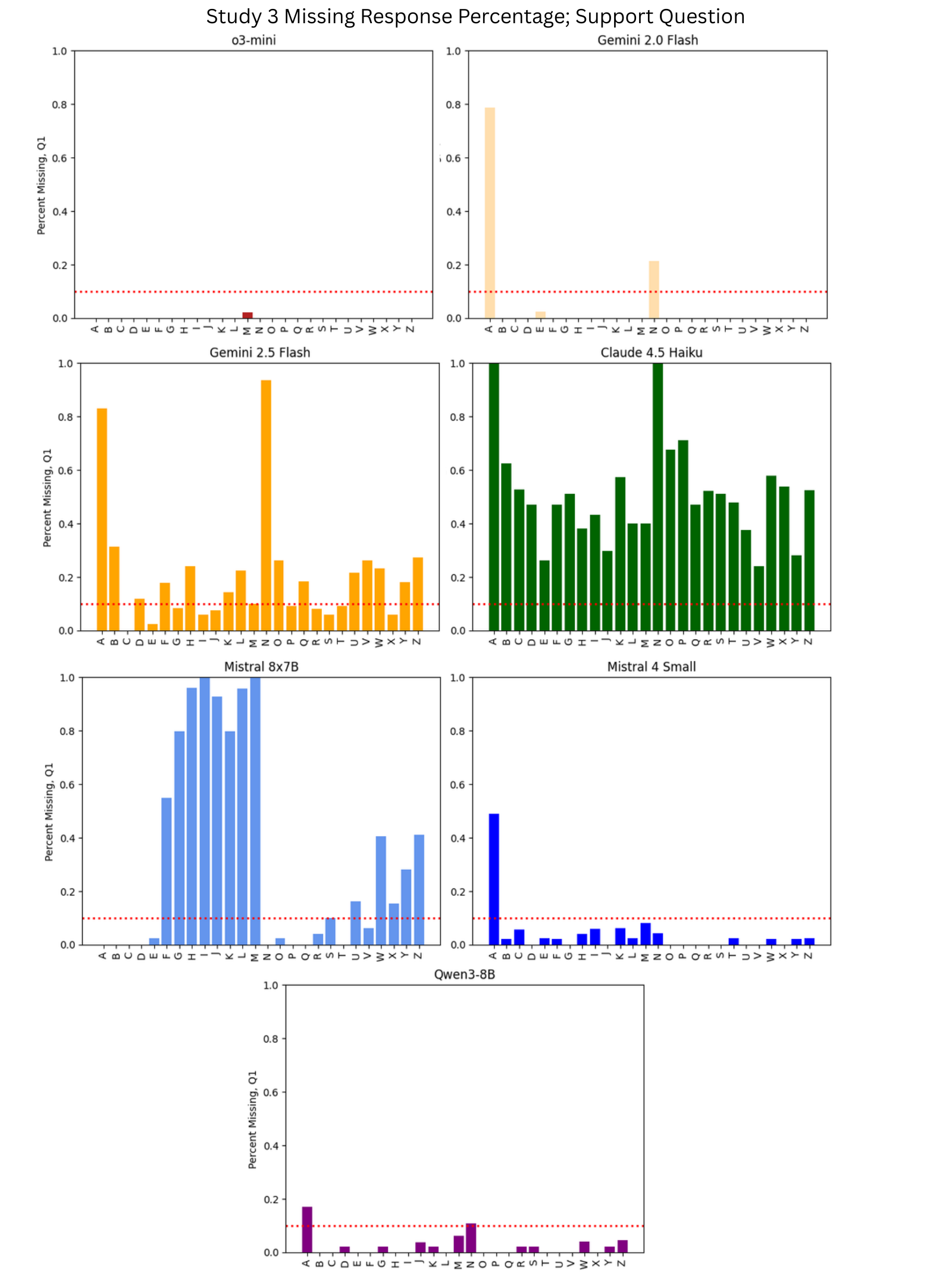}
    \caption{Percentage of missing data partitioned by model and condition in Study 3 for the questions asking for the model's support. 10\% denoted by red line.}
    \label{fig:study3_missing_support.pdf}
\end{figure*}

\begin{figure*}
    \centering
    \includegraphics[width=0.95\linewidth]{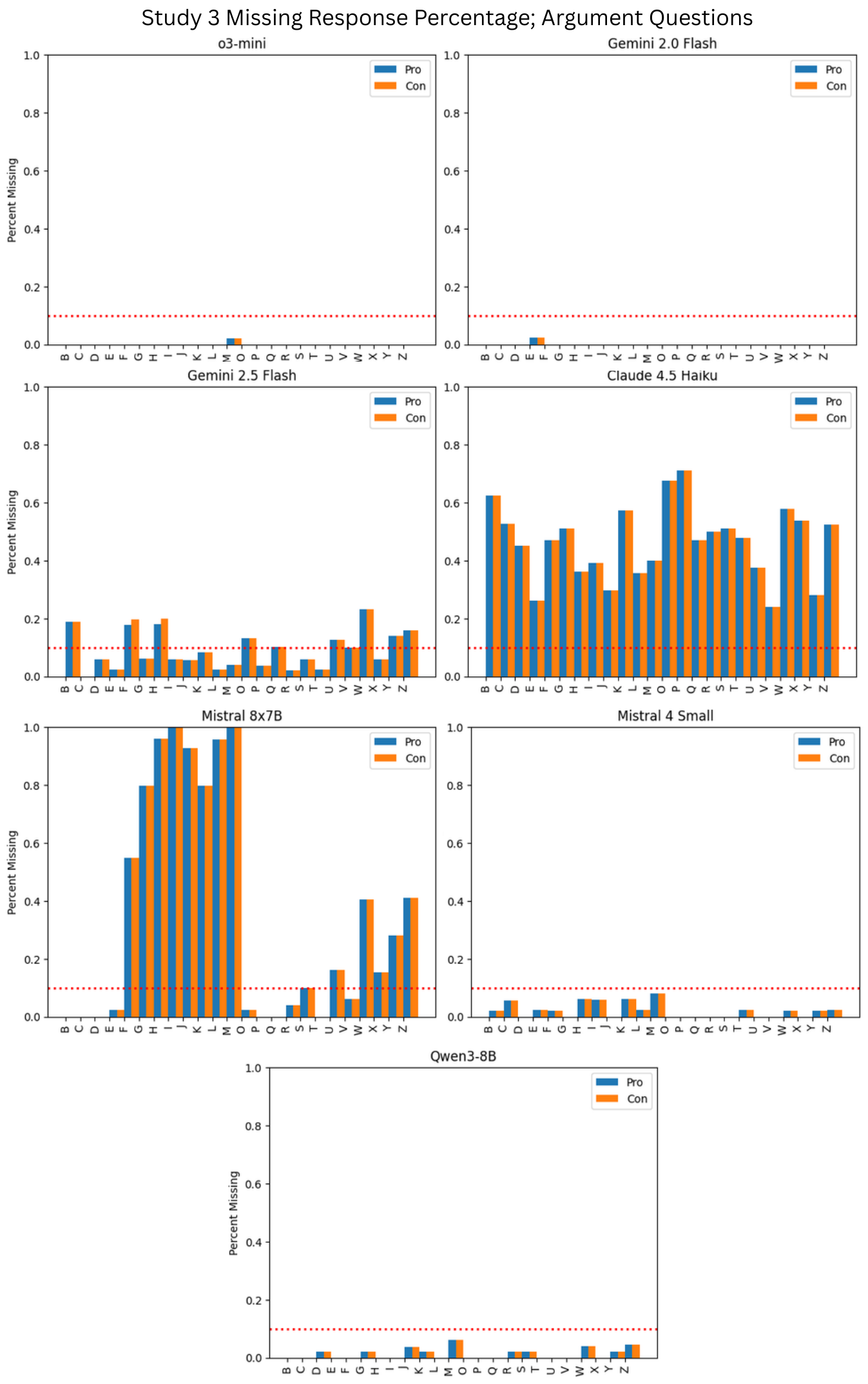}
    \caption{Percentage of missing data partitioned by model and condition in Study 3 for the questions asking for the model's argument assessment. 10\% denoted by red line.}
    \label{fig:study3_missing_arg.pdf}
\end{figure*}

\begin{figure*}
    \centering
    \includegraphics[width=0.95\linewidth]{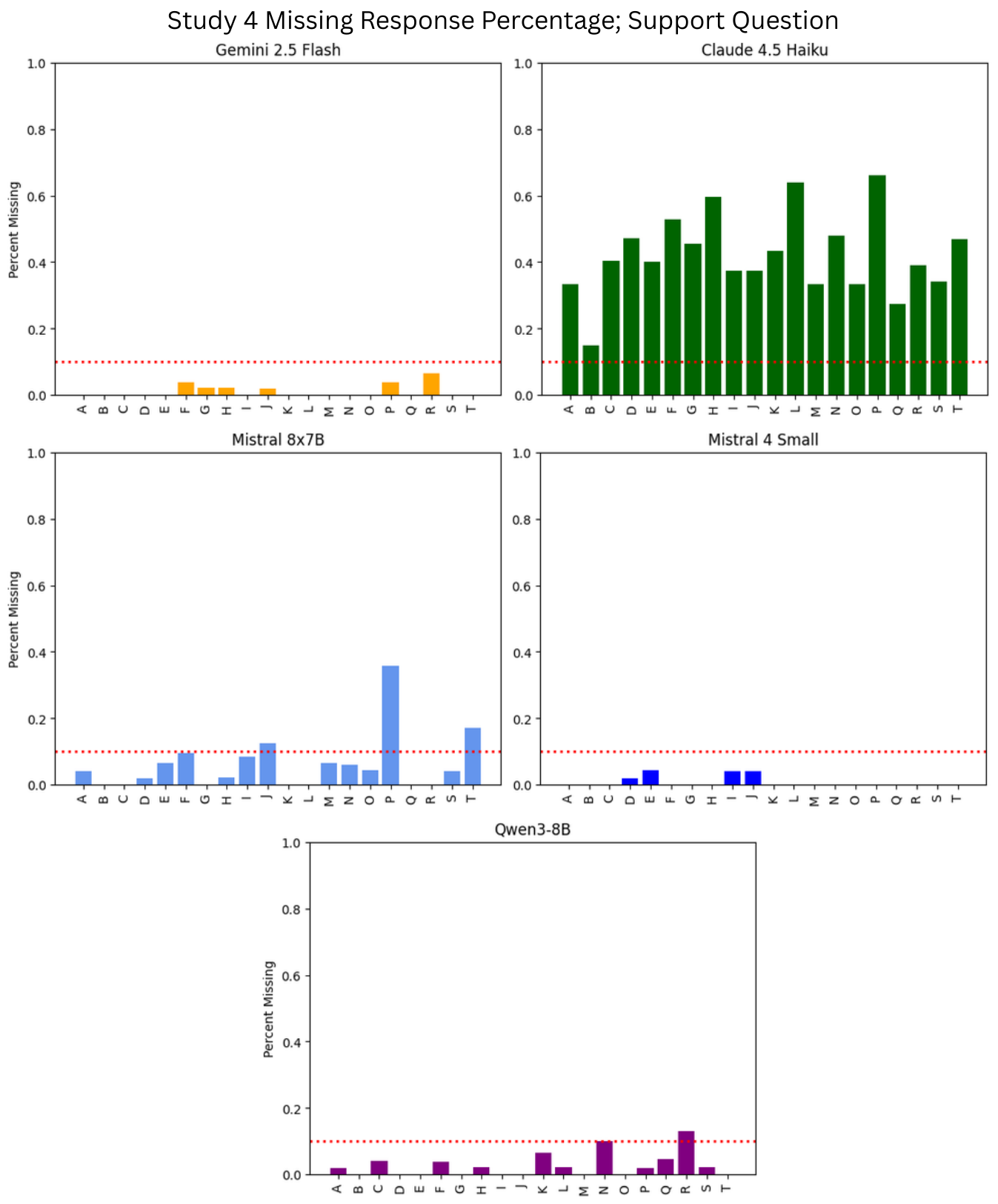}
    \caption{Percentage of missing data partitioned by model and condition in Study 4 for the questions asking for the model's support. 10\% denoted by red line.}
    \label{fig:study4_missing_support.pdf}
\end{figure*}

\begin{figure*}
    \centering
    \includegraphics[width=0.95\linewidth]{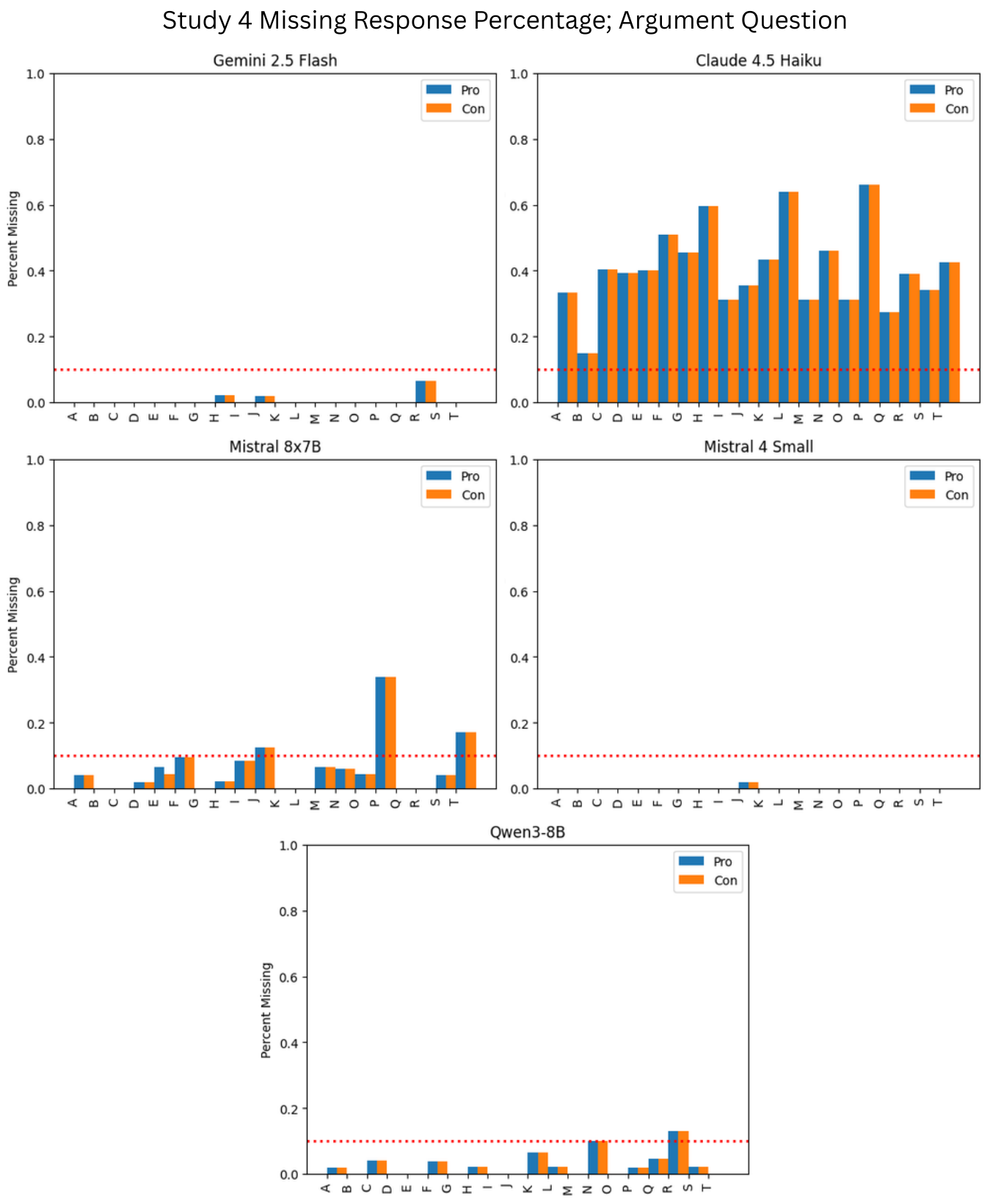}
    \caption{Percentage of missing data partitioned by model and condition in Study 4 for the questions asking for the model's argument assessment. 10\% denoted by red line.}
    \label{fig:study4_missing_arg.pdf}
\end{figure*}

\subsection{Cross LLM Behavior}
\label{sec:modelspecific}

\subsubsection{LLM Argument Strength Assessments}
\label{sec:argassessments}
We investigate further into how LLMs assess the strength of arguments when they are provided with them. In study 3, the LLM is provided with both pro and con arguments that vary in strength (either ``strong'' or ``weak'' as determined by the study). It stands to reason that if LLMs are not capable of motivated reasoning, then they should be able to accuracy assess which arguments are stronger than others. Comparing the difference in the pro argument and con argument strength across conditions (Figure \ref{fig:1.3_provcon_permodel.pdf}), we find that is not the case. While some unbiased assessments are in-tact in the ``Drilling'' conditions (e.g., if pro was strong and con was weak, we would expect a strong positive response) for o3-mini and Mistral 4 Small, this pattern is not maintained across the rest of the models nor in the ``DREAM'' condition of any model except Gemini 2.0 Flash.

In study 4, all arguments presented to the models are the same, where only the condition around the information is the same. However, we find that models are highly variable in their difference in pro and con argument assessments (Figure \ref{fig:1.4_provcon_permodel.pdf}) except for Gemini 2.5 Flash, Claude 3 Haiku, and Mistral 4 Small. Taken together, these results suggest that LLMs are not able to accurately assess the strength of arguments when providing information to them.

\subsubsection{Model Non-Response}
\label{sec:nonresponseanalysis}
When prompted to respond to the survey questions, different LLMs tend to ``opt-out'' based on different criteria included in the prompt. For example in Study 1, we see high amounts of non-response from Gemini 2.5 Flash and Claude 4.5 Haiku when ``no motivation'' is given, while Claude 3 Haiku fails to respond under only ``directional motivation'' conditions (Figure \ref{fig:study1_missing.pdf}). In study 2, we see high non-response from Gemini 2.0 Flash only in the ``Warning + Politicized scientific information, CNT'' condition; Gemini 2.5 Flash has high non-response rates under ``No information'' and also the majority of ``fracking'' based conditions, which differs from Claude 3 Haiku which has large non-response rates under ``CNT'' based conditions. Mistral 8x7B has lowest non-response rates under conditions which probe for accuracy. Only Claude 4.5 Haiku and Mistral 4 Small thought to be missing uniformly in this study (Figure \ref{fig:study2_missing.pdf}). Study 3 shows high non-response rates for Gemini 2.0 Flash, Gemini 2.5 Flash, Claude 4.5 Haiku, Mistral 8x7B, Mistral 4 Small, and Qwen3-8B under ``Control'' conditions. Mistral 8x7B experiences higher non-response rates when ``non-polarized parties'' and ``polarized parties'' are introduced (Figure \ref{fig:study3_missing_support.pdf}). These trends are largely reflected in the argument assessment non-response rates (Figure \ref{fig:study3_missing_arg.pdf}). There are no clear signals for which features cause opting on in Study 4, suggesting uniform non-response from all models (Figure \ref{fig:study4_missing_support.pdf}). These trends are largely reflected in the argument assessment non-response rates (Figure \ref{fig:study4_missing_arg.pdf}). Altogether, these results show how models will abstain differently based on the input information. Notably, the patterns in model opt-out seem to be study dependent (such as all models being affected by ``motivation'' in Study 1, all models being affected by ``Control'' in Study 3, and no clear signal in Study 4) rather than model family dependent (e.g., no clear patterns between Gemini 2.0 Flash and Gemini 2.5 Flash).

\begin{table*}[]
    \centering
    \begin{tabular}{|c|c|c|c|}
        \hline 
        \textbf{Model} & \textbf{Study} & \textbf{Chi-Square} & \textbf{Significant Partitions}\\ \hline
        Gemini 2.5 Flash & Study 1 & $73.952^{***}$ & Motivation \\ \hline
        Claude 3 Haiku &   & $156.665^{***}$ & Motivation \\ \hline
        Claude 4.5 Haiku &   & $73.819^{***}$ & Motivation \\ \hline \hline
        Gemini 2.0 Flash & Study 2 & $398.312^{***}$ & Technology, Information\\ \hline
        Gemini 2.5 Flash &   & $222.249^{***}$ & Information\\ \hline
        Claude 3 Haiku &   & $78.666^{***}$ & Technology \\ \hline
        Claude 4.5 Haiku &   & 7.604 & None \\ \hline
        Mistral 8x7B &   & $28.832^{**} $ & Information \\ \hline
        Mistral 4 Small &   & $29.01^{**} $ & None \\  \hline  \hline
        Gemini 2.0 Flash & Study 3 & $790.733^{***}$ & Control \\ \hline
        Gemini 2.5 Flash &   & $289.889^{***}$ & Control\\ \hline
        Claude 4.5 Haiku &  & $85.351^{***}$ & Control\\ \hline
        Mistral 8x7B &   & $579.157^{***}$ & Topic, Polarization, Control\\ \hline
        Mistral 4 Small &   & $291.9^{***}$ & Control \\ \hline 
        Qwen3-8B & & $85.848^{***}$ & Control \\ \hline \hline
        Claude 4.5 Haiku & Study 4 & $32.364^{*}$ & None \\ \hline
        Mistral 8x7B &  & $108.576^{***}$ & None \\ \hline
        Qwen3-8B & & $45.132^{***}$ & None \\ \hline
        
    \end{tabular}
    \caption{Chi-Square Results for Missing Responses Partitioned on Model and Study. Only considers models with at least 10\% of responses missing in at least one category. The ``Significant Partitions'' column represents which conditions the recomputed Chi-Square value was still significant after aggregating into each subcategory.}
    \label{tab:chisquare}
    \vspace{-1em}
\end{table*}

\subsubsection{Lack of Relationship in Behavior Between Reasoning/Non-Reasoning and Families}
Certain similarities in behavior in LLMs are domain dependent, model dependent, and even task dependent. For example, we see large groups of models are highly aligned in study 1 for change in opinion (Table \ref{tab:spearman1.1}) and study 3 for accuracy of argument assessment (Table \ref{tab:1.3_accuracy_provcon}). In some studies, we see relationships between certain models, such as study 2 for percent support (Table \ref{tab:spearman1.2}), study 4 for percent support (Table \ref{tab:spearman1.4}), study 1 for accuracy of the signs (Table \ref{tab:accuracy1.1}, and study 4 for accuracy of argument assessment (Table \ref{tab:1.4_accuracy_provcon}). All remaining comparisons (study 3 for percent change, Table \ref{tab:spearman1.3} and study 3 for accuracy of support, Table \ref{tab:accuracy1.3} show no relationships between any of the models.

Notably, this study include 5 model families, with two representatives from each family -- one reasoning and one non-reasoning. This gives us the ability to not only compare performance within families, but across thinking capabilities. While models within the same family sometimes share high correlations or accuracies, we only find consistency in the Mistral family across all studies where enough samples from both models were generated. This suggests a trend in misalignment across most families. Within model groups of reasoning and non-reasoning, we similarly find a lack of model alignment within the groups. Particularly for reasoning models, this may suggest that models are reasoning about their responses differently, leading to different measured outcomes. This is important to keep in mind when selecting which model to use for study replications.

\begin{table*}
\centering
\begin{tabular}{|l|p{0.1\textwidth}|p{0.1\textwidth}|p{0.1\textwidth}|p{0.1\textwidth}|p{0.1\textwidth}|p{0.1\textwidth}|p{0.1\textwidth}|}
\hline 
 & o3-mini & Gemini 2.0 Flash & Mistral 8x7B & Mistral 4 Small & Qwen2.5-7B-Instruct & Qwen3-8B\\ \hline
GPT 4o-mini &  \cellcolor{green!30}$0.746^{***}$ & \cellcolor{gray!30}0.191 & \cellcolor{green!30}$0.808^{***}$ & \cellcolor{green!30}$0.673^{**}$ & \cellcolor{yellow!30}$0.499^{*}$ &\cellcolor{yellow!30} $0.577^{**}$\\
o3-mini & &  \cellcolor{gray!30}0.262 & \cellcolor{green!30}$0.617^{**}$ & \cellcolor{green!30}$0.736 ^{***}$& \cellcolor{yellow!30}$0.493^{*}$ & \cellcolor{green!30}$0.692^{***}$\\
Gemini 2.0 Flash & &  & \cellcolor{gray!30}0.096 & \cellcolor{gray!30}0.087 & \cellcolor{yellow!30}0.531 & \cellcolor{gray!30}0.378 \\
Mistral 8x7B &  &   & & \cellcolor{green!30}$0.714^{***}$ & \cellcolor{gray!30}0.362 & \cellcolor{yellow!30}$0.448^{*}$\\
Mistral 4 Small &  &  &   & & \cellcolor{green!30}0.237 & \cellcolor{yellow!30}$0.473^{*}$ \\ 
Qwen2.5-7B-Instruct & & & & & & \cellcolor{yellow!30}$0.566^{*}$\\ \hline
\end{tabular}

  \caption{Kendall's tau measures between average percent change for each condition across all models for study 1. * = p value \textless 0.05, ** = p value \textless 0.01, *** = p value \textless 0.001. Gray signifies weak positive correlation, yellow signifies moderate positive correlation, and green signifies strong positive correlation. Models that could not generate enough samples are dropped.}
  \label{tab:spearman1.1}
\end{table*}

\begin{table*}
\centering
\begin{tabular}{|l|p{0.1\textwidth}|p{0.1\textwidth}|p{0.1\textwidth}|p{0.1\textwidth}|p{0.1\textwidth}|p{0.1\textwidth}|p{0.1\textwidth}|p{0.1\textwidth}|}
\hline 
 & o3-mini & Claude 3 Haiku & Qwen2.5-7B-Instruct & Qwen3-8B\\ \hline
GPT 4o-mini & \cellcolor{gray!30}0.283 & \cellcolor{green!30}$0.683^{*}$ & \cellcolor{green!30}$0.724^{*}$ & \cellcolor{gray!30}0.381 \\
o3-mini &   & \cellcolor{gray!30}0.314 & \cellcolor{yellow!30}0.538 & \cellcolor{yellow!30}0.531\\
Claude 3 Haiku &  &  & \cellcolor{yellow!30}0.494 & \cellcolor{gray!30}0.163 \\ 
Qwen2.5-7B-Instruct & & & & \cellcolor{yellow!30}$0.592^{*}$ \\ \hline

\end{tabular}
  \caption{Kendall's tau measure between  percent support for each condition across all models for study 2. * = p value \textless 0.05, ** = p value \textless 0.01, *** = p value \textless 0.001. Gray signifies weak positive correlation, yellow signifies moderate positive correlation, and green signifies strong positive correlation.}
  \label{tab:spearman1.2}
\end{table*}

\begin{table*}
\centering
\begin{tabular}{|l|p{0.1\textwidth}|p{0.1\textwidth}|p{0.1\textwidth}|p{0.1\textwidth}|p{0.1\textwidth}|p{0.1\textwidth}|p{0.1\textwidth}|p{0.1\textwidth}|}
\hline 
 & o3-mini & Claude 3 Haiku & Qwen2.5-7B-Instruct \\ \hline
GPT 4o-mini  & \cellcolor{red!30}-0.152 & \cellcolor{yellow!30}$0.565^{***}$ & \cellcolor{red!30}-0.23 \\
o3-mini & & \cellcolor{red!30}-0.239 & \cellcolor{gray!30}0.305 \\
Claude 3 Haiku & & & \cellcolor{red!30}-0.088\\ \hline 

\end{tabular}
  \caption{Kendall's tau between percent change in response for each condition across all models for study 3. * = p value \textless 0.05, ** = p value \textless 0.01, *** = p value \textless 0.001. Red signifies negative correlation, gray signifies weak positive correlation, and green signifies strong positive correlation.}
  \label{tab:spearman1.3}
\end{table*}

\begin{table*}
\centering
\begin{tabular}{|l|p{0.1\textwidth}|p{0.1\textwidth}|p{0.1\textwidth}|p{0.1\textwidth}|p{0.1\textwidth}|p{0.1\textwidth}|p{0.1\textwidth}|p{0.1\textwidth}|}
\hline 
 & o3-mini & Gemini 2.0 Flash & Claude 3 Haiku & Mistral 4 Small & Qwen2.5-7B-Instruct \\
\hline
GPT 4o-mini & \cellcolor{yellow!30}$0.517^{**}$ & \cellcolor{gray!30}0.067 & \cellcolor{gray!30}0.184 & \cellcolor{yellow!30}$0.467^{**}$ & \cellcolor{yellow!30}$0.568^{**}$ \\
o3-mini & & \cellcolor{yellow!30}$0.511^{**}$ & \cellcolor{green!30}$0.669^{***}$ & \cellcolor{green!30}$0.727^{***}$ & \cellcolor{green!30}$0.617^{***}$ \\
Gemini 2.0 Flash & & & \cellcolor{yellow!30}$0.593^{**}$ & \cellcolor{yellow!30}$0.562^{**}$ & \cellcolor{yellow!30}$0.537^{**}$ \\
Claude 3 Haiku & & & & \cellcolor{yellow!30}$0.488^{**}$ & \cellcolor{gray!30}$0.358^{*}$ \\
Mistral 4 Small & & & & & \cellcolor{green!30}$0.693^{***}$\\ \hline

\end{tabular}
  \caption{Kendall's tau measure between average response for each condition across all models for study 4. * = p value \textless 0.05, ** = p value \textless 0.01, *** = p value \textless 0.001. Gray signifies weak positive correlation and green signifies strong positive correlation.}
  \label{tab:spearman1.4}
\end{table*}

\begin{table*}
\centering
\begin{tabular}{|l|p{0.1\textwidth}|p{0.1\textwidth}|p{0.1\textwidth}|p{0.1\textwidth}|p{0.1\textwidth}|p{0.1\textwidth}|p{0.1\textwidth}|p{0.1\textwidth}|}
\hline 
 & o3-mini & Gemini 2.0 Flash & Mistral 8x7B & Mistral 4 Small & Qwen2.5-7B-Instruct & Qwen3-8B\\ \hline
 
GPT 4o-mini &  \cellcolor{gray!30}0.429 &  \cellcolor{green!30}0.714 &  \cellcolor{yellow!30}0.643 &  \cellcolor{yellow!30}0.5 &  \cellcolor{gray!30}0.286 &  \cellcolor{yellow!30}0.643\\
o3-mini & &   \cellcolor{gray!30}0.143 &  \cellcolor{yellow!30}0.643 &  \cellcolor{green!30}0.786 &  \cellcolor{green!30}0.786 &  \cellcolor{yellow!30}0.571\\
Gemini 2.0 Flash & &  &  \cellcolor{gray!30}0.357 &  \cellcolor{gray!30}0.214 & 0 &  \cellcolor{gray!30}0.357\\
Mistral 8x7B &  &   & &  \cellcolor{green!30}0.857 &  \cellcolor{yellow!30}0.571 &  \cellcolor{yellow!30}0.571 \\
Mistral 4 Small &  &  &   & &  \cellcolor{yellow!30}0.643 & \cellcolor{yellow!30}0.571 \\ 
Qwen2.5-7B-Instruct & & & & & & \cellcolor{yellow!30}0.571 \\ \hline

\end{tabular}
  \caption{Accuracy of expected signs for each condition across pairs of models for study 1. Gray signifies low accuracy, yellow signifies moderate accuracy, and green signifies high accuracy.}
  \label{tab:accuracy1.1}
\end{table*}

\begin{table*}
\centering
\begin{tabular}{|l|p{0.1\textwidth}|p{0.1\textwidth}|p{0.1\textwidth}|p{0.1\textwidth}|p{0.1\textwidth}|p{0.1\textwidth}|p{0.1\textwidth}|}
\hline 
 & o3-mini & Claude 3 Haiku & Qwen2.5-7B-Instruct \\ \hline
GPT 4o-mini  & \cellcolor{gray!30}0.083 & \cellcolor{yellow!30}0.583  & \cellcolor{gray!30}0.125 \\
o3-mini & & \cellcolor{gray!30}0.292 & \cellcolor{gray!30}0.208 \\
Claude 3 Haiku & & & \cellcolor{gray!30}0.208\\ \hline 
\end{tabular}
  \caption{Accuracy of expected sign of change in support for each condition across pairs of models for study 3. Gray signifies low accuracy, and yellow signifies moderate accuracy.}
  \label{tab:accuracy1.3}
\end{table*}

\begin{table*}
\centering
\begin{tabular}{|l|p{0.1\textwidth}|p{0.1\textwidth}|p{0.1\textwidth}|p{0.1\textwidth}|p{0.1\textwidth}|p{0.1\textwidth}|p{0.1\textwidth}|p{0.1\textwidth}|p{0.1\textwidth}|p{0.1\textwidth}|}
\hline 
 & o3-mini & Gemini 2.0 Flash & Claude 3 Haiku & Mistral 8x7B & Mistral 4 Small & Qwen2.5-7B-Instruct & Qwen3-8B \\ \hline
GPT-4o mini & \cellcolor{green!30}0.791 & \cellcolor{gray!30}0.292 & \cellcolor{green!30}0.875 & \cellcolor{yellow!30}0.583 & \cellcolor{green!30}0.708 & \cellcolor{yellow!30}0.667 & \cellcolor{yellow!30}0.542\\ 
o3-mini & & \cellcolor{gray!30}0.25 & \cellcolor{green!30}0.75 & \cellcolor{green!30}0.75 & \cellcolor{green!30}0.917 & \cellcolor{green!30}0.75 & \cellcolor{green!30}0.708\\ 
Gemini 2.0 Flash & & & \cellcolor{gray!30}0.292 & \cellcolor{gray!30}0.417 & \cellcolor{gray!30}0.292 & \cellcolor{gray!30}0.333 & \cellcolor{gray!30}0.458\\ 
Claude 3 Haiku & & & & \cellcolor{yellow!30}0.583 & \cellcolor{green!30}0.75 & \cellcolor{yellow!30}0.625 & \cellcolor{yellow!30}0.5 \\ 
Mistral 8x7B & & & & & \cellcolor{green!30}0.75 & \cellcolor{yellow!30}0.667 & \cellcolor{green!30}0.875\\ 
Mistral 4 Small & & & & & & \cellcolor{yellow!30}0.667 & \cellcolor{green!30}0.708\\
Qwen2.5-7B-Instruct & & & & & & & \cellcolor{green!30}0.708 \\
\hline 

\end{tabular}
  \caption{Accuracy of expected sign of pro - con for each condition across pairs of models for study 3. Yellow signifies moderate accuracy, and green signifies high accuracy.}
  \label{tab:1.3_accuracy_provcon}
\end{table*}

\begin{table*}
\centering
\begin{tabular}{|l|p{0.1\textwidth}|p{0.1\textwidth}|p{0.1\textwidth}|p{0.1\textwidth}|p{0.1\textwidth}|p{0.1\textwidth}|p{0.1\textwidth}|p{0.1\textwidth}|}
\hline 
 & o3-mini & Gemini 2.0 Flash & Gemini 2.5 Flash & Claude 3 Haiku & Mistral 4 Small & Qwen2.5-7B-Instruct \\  \hline
GPT-4o mini & \cellcolor{gray!30}0.35 & \cellcolor{yellow!30}0.5 & \cellcolor{gray!30}0.15 & \cellcolor{green!30}0.7 & \cellcolor{yellow!30}0.55 & \cellcolor{gray!30}0.3 \\ 
o3-mini & & \cellcolor{yellow!30}0.5 & \cellcolor{gray!30}0.4 & \cellcolor{gray!30}0.3 & \cellcolor{gray!30}0.45 & \cellcolor{yellow!30}0.5 \\ 
Gemini 2.0 Flash & & & \cellcolor{gray!30}0.15 & \cellcolor{yellow!30}0.65 & \cellcolor{yellow!30}0.65 & \cellcolor{gray!30}0.2 \\
Gemini 2.5 Flash & & & & \cellcolor{gray!30}0.25 & \cellcolor{gray!30}0.3 & \cellcolor{gray!30}0.35 \\ 
Claude 3 Haiku & & & & & \cellcolor{green!30}0.75 & \cellcolor{gray!30}0.15 \\
Mistral 4 Small & & & & & & \cellcolor{gray!30}0.25 \\
\hline 

\end{tabular}
  \caption{Accuracy of expected sign of pro - con for each condition across pairs of models for study 4. Gray signifies low accuracy, yellow signifies moderate accuracy, and green signifies high accuracy.}
  \label{tab:1.4_accuracy_provcon}
\end{table*}

\subsection{Additional Condition: Prompting for Reasoning}
\label{sec:justification}
As an additional measure for this study, we replace the prompt directive asking for <question number>:<numeric answer> with ``please provide your reasoning before the final answer'' to the end of all LLM prompts on a subset of models. We focus on the set of non-reasoning models, since reasoning models are likely already going through deliberation behind the scenes. This adaptation was tested for several reasons. First, this is meant to stimulate internally-accurate answers, similar to a chain-of-thought prompt \cite{10.5555/3600270.3602070}, and minimize hallucinations. This may also help standardize LLM responses, as some LLMs may be more verbose and inherently provide reasoning without being asked to do so or may refrain from responding without being asked for reasoning. While this may seem similar to an accuracy motivation by asking the LLM to provide its reasoning, it is important to differentiate the difference between ``justification'' and ``communication''. Since justification entails having to defend a given position, this differs from communication without defense (simply providing reasoning) \cite{Tetlock_1983} and thus does not invoke an accuracy motivation. This additionally hoped to solve some of the non-response problems from the main paper.

Because this response was more free-form (because of the added reasoning), all responses were first cleaned to remove repetitions of the numeric scale the LLM was asked to stay within (``1 to 7'' or ``0 to 10''). Such examples of parsing include (1) ``1-7'', ``1-7'', ``1 to 7'', ``0-10'' as a repetition of scale, (2)  `/7'', ``out of 7'', ``/10'', ``out of 10'', signifying a score out of 7 or 10 (such as 5/7 or 6/10), and (3) ``where 1 = ... and 7 = ...'' as repetition of the meaning of the bounds. Only numeric answers were counted; ``1'' was accepted while ``one'' was not. This is because the initial studies ask for a numeric answer and the scale is provided to the LLMs as ``1-7'' and ``0-10''. 

When numeric answers could not be detected for a question response, the given generation number and condition was flagged. A researcher went in and determined whether there was no response or if there was a response, reformatted so the numeric answer could be detected. If no numeric was found, the question was dropped.

We replicate the main paper findings only. We find similar trends of moderate / low / negative correlations in Table \ref{tab:humanvllm_provide}, moderate accuracies in Table \ref{tab:accuracysign_provide}, and moderate / low correlations and accuracies in Table \ref{tab:spearman_meanvprovcon_provide}. Asking for justification does improve alignment between the model's support and their pro - con assessments (Table \ref{tab:spearman_meanvprovcon_provide}), which may suggest asking the LLM to provide reasoning causes incorporation of the information into its assessment. Notably, most correlations are still lower than the human baselines.

Remarkably, prompting for providing reasoning also causes the LLMs to provide more answers. There were no LLMs with missing responses below 10\% in Study 1 and Study 4. Claude 3 Haiku missed 8 responses to ``No information, CNT'' in Study 2. GPT-4o mini missed 18 on ``Control, DREAM'' and Claude 3 Haiku missed 7 on ``Control, Drilling'' for question 1 only in Study 3, leaving them with enough responses to analyze their response to the pro / con arguments. Gemini 2.0 Flash missed 8 for ``High polarization, low importance, traditional endorsement, tax'' and 5 for ``high polarization, high importance, reversed endorsement, education'' for questions 2 and 3 in Study 4, providing enough responses for analysis regarding question 1 (the support question). Regardless, the addition of providing reasoning does not seem to improve alignment between human and LLM responses.


\begin{table*}[ht]
  \centering
  \begin{tabular}{|p{0.25\linewidth}|p{0.08\linewidth}|p{0.08\linewidth}|p{0.1\linewidth}|p{0.08\linewidth}||p{0.21\linewidth}|}
    \hline
    \textbf{Model} & \textbf{Study 1} & \textbf{Study 2} & \textbf{Study 3} & \textbf{Study 4} & \textbf{Average Correlation}\\
    \hline
    \textbf{GPT-4o mini} & \cellcolor{gray!30}0.384 &  \cellcolor{gray!30}0.094 & N/A  & \cellcolor{gray!30}0.147 & \cellcolor{gray!30}0.208\\
    \textbf{Gemini 2.0 Flash} & \cellcolor{gray!30}0.291 &  \cellcolor{gray!30}0.021 &  \cellcolor{red!30}$-0.41^{**}$& \cellcolor{gray!30}0.026 & \cellcolor{gray!30}-0.018\\
    \textbf{Claude 3 Haiku} & \cellcolor{gray!30}0.121 &  N/A & N/A & \cellcolor{gray!30}0.032 & \cellcolor{gray!30}0.077\\ 
    \textbf{Mistral 8x7B} & \cellcolor{gray!30}0.376 &  \cellcolor{gray!30}0.333& \cellcolor{red!30}-0.16 & \cellcolor{red!30}-0.021 & \cellcolor{gray!30}0.132\\
    \textbf{Qwen2.5-7B-Instruct} & \cellcolor{gray!30}0.223 & \cellcolor{gray!30} 0.305 &  \cellcolor{red!30}-0.16 & \cellcolor{red!30}-0.053 & \cellcolor{gray!30}0.079\\ \hline
  \end{tabular}
  \caption{Replication of Table 1, ``Correlations Between Human and LLM Averages'', on Non-Reasoning Models Prompted to Provide Reasonings. Correlations are computed with Kendall's Tau on the reported average (Study 4), change in average compared to control (Studies 1 and 3), or reported overall support (Study 2). N/A signifies not enough samples were generated in at least one condition to make a fair comparison. * = p value \textless 0.05. Red signifies negative relationships, gray signifies weak positive relationships, and yellow signifies moderate positive relationships.}
  \label{tab:humanvllm_provide}
\end{table*}

\begin{table}[ht]
  \centering
  \begin{tabular}{|p{0.48\linewidth}|p{0.2\linewidth}|p{0.2\linewidth}|}
    \hline
    \textbf{Model} & \textbf{Study 1} & \textbf{Study 3} \\
    \hline
    \textbf{GPT-4o mini} & \cellcolor{yellow!30}0.571 & N/A  \\
    \textbf{Gemini 2.0 Flash} & \cellcolor{green!30}0.714 &  \cellcolor{yellow!30}0.167 \\
    \textbf{Claude 3 Haiku} & \cellcolor{yellow!30}0.5 &  N/A \\
   \textbf{Mistral 8x7B} & \cellcolor{gray!30}0.428 & \cellcolor{yellow!30}0.167 \\
    \textbf{Qwen2.5-7B-Instruct} & \cellcolor{yellow!30}0.5 & \cellcolor{yellow!30}0.333 \\
    \hline
  \end{tabular}
  \caption{Replication of Table 2, ``Accuracy of the Sign of the Change in Opinion Relative to the Control'', on Non-Reasoning Models Prompted to Provide Reasoning. When the change compared to control is positive and significant, this is encoded as 1. When the change compared to the control is negative and significant, this is encoded as -1. All other cases are encoded as 0. N/A signifies not enough samples were generated in at least one condition to make a fair comparison. Gray signifies weak accuracy, yellow signifies moderate accuracy, and green signifies high accuracy.}
  \label{tab:accuracysign_provide}
\end{table}

\begin{table*}
  \centering
  \begin{tabular}{|c|c|c||c|c|}
    \hline
    \textbf{Model} & \textbf{Study 3 Avg.} & \textbf{Study 4 Avg.} & \textbf{Study 3 Sign} & \textbf{Study 4 Sign} \\
    \hline
    \textbf{GPT-4o mini} &\cellcolor{yellow!30}$0.415^{**}$ &  \cellcolor{gray!30}0.195 & \cellcolor{yellow!30}0.5 &  \cellcolor{gray!30}0.4 \\
    \textbf{Gemini 2.0 Flash} & \cellcolor{gray!30}0.237 & N/A & \cellcolor{yellow!30}0.5 & N/A \\
    \textbf{Claude 3 Haiku} & \cellcolor{yellow!30}$0.472^{**}$ & \cellcolor{gray!30}0.317 & \cellcolor{yellow!30}0.5 & \cellcolor{gray!30}0.45 \\
    \textbf{Mistral 8x7B} & \cellcolor{yellow!30}$0.465^{**}$ & \cellcolor{gray!30}0.1  & \cellcolor{yellow!30}0.5 &  \cellcolor{gray!30}0.35 \\
    \textbf{Qwen2.5-7B-Instruct} & \cellcolor{yellow!30}$0.487^{***}$ &  \cellcolor{gray!30}0.058 & \cellcolor{gray!30}0.458 & \cellcolor{gray!30}0.35  \\ \hline
  \end{tabular}
  \caption{Replication of Table 3, ``Correlation and Accuracy Between Human and LLM Argument Evaluation'', on Non-Reasoning Models Prompted to Provide Reasoning. ``Avg.'' correlations are computed with Kendall's tau on the reported average pro - average con message. ``Sign'' looks at the accuracy of the expected sign for each condition. When the change compared to control is positive and significant, this is encoded as 1. When the change compared to the control is negative and significant, this is encoded as -1. All other cases are encoded as 0. * = p value \textless 0.05, ** = p value \textless 0.01, *** = p value \textless 0.001. For ``Avg.'', red signifies negative relationships, gray signifies weak positive relationships, and yellow signifies moderate positive relationships. For ``Sign'', gray signifies low accuracy.}
  \label{tab:provcon_provide}

\end{table*}

\begin{table}[ht]
\centering
\begin{tabular}
{|l|p{0.1\textwidth}|p{0.1\textwidth}|p{0.1\textwidth}|}
\hline 
 \textbf{Model} & \textbf{Study 3} & \textbf{Study 4} \\
\hline 
\textbf{GPT-4o mini} & \cellcolor{green!30}$0.79^{***}$& \cellcolor{green!30}$0.652^{***}$\\
\textbf{Gemini 2.0 Flash} & \cellcolor{green!30}$0.606^{***}$& N/A  \\
\textbf{Claude 3 Haiku} & \cellcolor{green!30}$0.737^{***}$& \cellcolor{green!30}$0.66^{***}$\\
\textbf{Mistral 8x7B} & \cellcolor{green!30}$0.717^{***}$& \cellcolor{gray!30}0.27\\
\textbf{Qwen2.5-7B-Instruct} & \cellcolor{yellow!30}$0.446^{**}$& \cellcolor{green!30}$0.684^{***}$\\
\hline \hline
\textbf{Human Baseline} & \cellcolor{green!30}$0.788^{***}$ & \cellcolor{green!30}$0.904^{***}$ \\
\hline 

\end{tabular}
  \caption{Replication of Table 4, ``Kendall's tau of the Average Opinion in Each Condition and the  Assessment of Argument Strength (Pro - Con) in Each Condition'', on Non-Reasoning Models Prompted to Provide Reasoning. N/A signifies not enough samples were generated in at least one condition to make a fair comparison. * = p value \textless 0.05, ** = p value \textless 0.01, *** = p value \textless 0.001. Gray signifies weak positive relationships, yellow signifies moderate positive relationships, and green signifies strong positive relationships.}
  \label{tab:spearman_meanvprovcon_provide}

\end{table}

\subsection{Probing LLMs with Demographics}
\label{sec:demographictest}

In addition to testing whether adding reasoning improves performance, we also explore whether providing the LLMs with demographics improves performance, since reflecting the distribution of individuals in the study originally may change the distribution of model answers. For this analysis, we rerun Study 2, since Study 1 and Study 3 do not provide demographic distributions of their participants and Study 4 provides the raw data. Study 2 also offers opportunity for the most improvement, with all original findings being poor (Table \ref{tab:humanvllm}) and models (almost) unanimously expressing support in all conditions (Figure \ref{fig:1.2_percentchangepercondition.pdf}). Additionally, Study 2 has successful analysis for the reasoning and non-reasoning models in two families in the main paper, allowing us to compare performance across families and reasoning capabilities.

For replicating the demographic distribution from the original study, we use the sampling method from prior work \cite{Westwood_2025} to generate personas for the LLMs from the original study distribution. Reasoning model token limits were raised to 4000 to allow for further reasoning of alignment based on persona inputs. We include all demographics originally collected except for questions that lacked a full mapping of numbers to text descriptions and questions where the probability distribution could not be fully recovered (e.g., correctness on a set of questions is provided in aggregate rather than by question). As such, the following demographics were given to the LLM:

\begin{itemize}
    \item Gender
    \item Race
    \item Age
    \item Education
    \item Party Affiliation / strength of affiliation
    \item Income level
    \item Ideology
    \item Interest in politics
    \item Discussion of politics
    \item Interest in energy politics
\end{itemize}

\begin{figure*}
    \centering
    \includegraphics[width=0.95\linewidth]{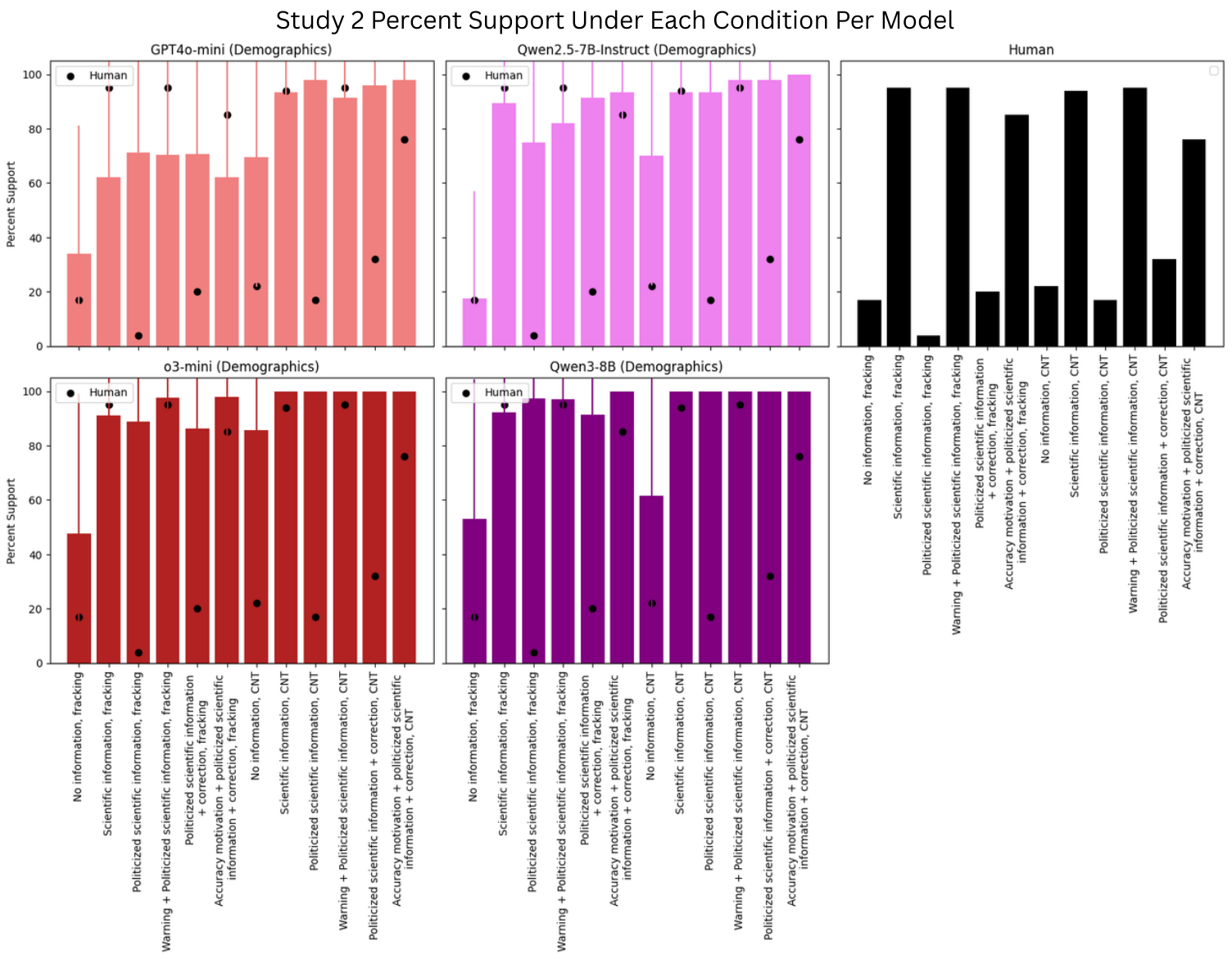}
    \caption{Replication of Figure 3, ``Results for the average percent of support, partitioned by model and condition, compared against the
reported human percent change in support (Study 2)''. Human results from the original paper provided in black for
reader convenience.}
    \label{fig:1.2_ablation_demo}
\end{figure*}

While the personality inductions do change the behavior of the LLM responses (Figure \ref{fig:1.2_ablation_demo}) compared to the outcomes of the original study, we see these trends clearly do not reflect the human trends from this study. This is supported by low, non-significant Kendall Tau correlation measures (all $\tau$ \textless 0.24). This supports the robustness of the main paper results.

\end{document}